\documentclass[a4paper,11pt]{article}
\usepackage{color}
\usepackage{jheppub} 
\usepackage{lineno}
\usepackage{macros}
\usepackage{stmaryrd}
\usepackage{theorems}
\usepackage{amsthm}

\title{Elementary Constituents Conjecture}

\author[a]{Vinicius Nevoa}
\author[a]{Sanjay Raman}
\author[a]{Cumrun Vafa}

\emailAdd{vnevoa@g.harvard.edu}
\emailAdd{sanjayraman@g.harvard.edu}
\emailAdd{vafa@g.harvard.edu}

\affiliation[a]{Jefferson Physical Laboratory, Harvard University\\ 17 Oxford St, Cambridge, MA 02138, United States of America} 

\abstract{In this note, we conjecture and provide evidence that quantum gravity cobordism classes are trivialized by elementary generators with tension $\lesssim 1$ in Planck units. Motivated by the attractor mechanism and a number of key examples, we propose that there is always a regime in the field space of the EFT in which the intrinsic tension of such an object is at most Planckian. This work brings together kinematic and dynamical aspects of cobordism defects in the context of the Swampland program.}

\begin{document}
\maketitle
\flushbottom

\section{Introduction}
The triviality of cobordism classes in quantum gravity proposed in \cite{mcnamara_cobordism_2019} has been tested extensively \cite{debray_chronicles_2023, montero_cobordism_2021, mcnamara_gravitational_2021, dierigl_swampland_2021, buratti_dynamical_2021} and has become one of the firm principles of the Swampland program \cite{van_beest_lectures_2022, agmon_lectures_2023}.  This conjecture, which is motivated by the absence of global symmetries in quantum gravity, leads to powerful implications for quantum gravitational theories. In particular, the Cobordism Conjecture (CC) has predicted many new objects (including non-supersymmetric ones) which are needed to trivialize various bordism classes. Compare this also with the predictions of the original No-Global-Symmetries Conjecture \cite{banks_symmetries_2011, harlow_symmetries_2019}; for example, the absence of 1-form global symmetries implies the completeness of electric charges in $U(1)$ gauge theories \cite{rudelius_topological_2020, heidenreich_non-invertible_2021}.

However, in the case of $U(1)$ gauge theories, there are further Swampland criteria which additionally put a bound on the masses of charged states predicted by the CC. In particular, the Weak Gravity Conjecture (WGC) \cite{arkani-hamed_string_2007,harlow_weak_2023} predicts that the masses of some elementary charged states are bounded by their charges, relative to the charge-to-mass ratio of an extremal black hole. It is natural to ask if {\it all} objects predicted by the CC satisfy such bounds --- for example, there is apparently no Swampland constraint providing an upper bound on the mass of states carrying $\B{Z}_n$ discrete gauge charges. It is the aim of this paper to present evidence that the generators of bordism groups can be trivialized by objects which have a mass bounded above by the Planck scale. (We sometimes loosely refer to these objects as ``bordism generators'' or as ``carrying bordism charge'' even though it is more precise to say that the class of a bordism group generator is trivialized by them.) This idea reinforces the picture that all objects are made of elementary constituents with minimal mass/tension.  In the case of theories with massless moduli, as in supergravity theories, the statement is that there exist regions in the moduli space, which could in principle depend on the object of interest, where the mass/tension of basic constituents is $\lesssim 1$ in Planck units. These points in the moduli space are determined by solving the field equations near the basic objects, (i.e. they are fixed by an ``attractor flow'' mechanism).

The organization of this paper is as follows. In Sec. \ref{sec:topological-charges-bordism}, we give a brief overview of the CC and its implications. In Sec. \ref{sec:ecc-motivations} we present our conjecture and give some heuristic motivation for it.  In Sec. \ref{sec:attractor} we discuss attractor flows and their relation to our conjecture. In Secs. \ref{sec:susy-examples} and \ref{ssec:non-susy-heterotic-type-i}, we discuss some examples of the conjecture, highlighting several non-supersymmetric tests of our conjecture in particular in Sec. \ref{ssec:non-susy-heterotic-type-i}. We move towards possible applications in phenomenology in Sec. \ref{sec:applications-pheno}, and we end with some concluding thoughts in Sec. \ref{sec:discussion}.

\section{Topological Charges and Bordism} \label{sec:topological-charges-bordism}

In this section, we review some necessary facts about topological charges in quantum field theory and quantum gravity as well as their relationship with the Cobordism Conjecture (CC). We review the original formulation of the CC in Sec. \ref{ssec:review-cobordism-conjecture}, elaborating in particular on the case of \textit{gauging} a would-be global symmetry associated to a nontrivial bordism group in Sec. \ref{ssec:gauged-bordism-defects}. Finally, in Sec. \ref{ssec:duality-topological-charges}, we argue that bordism is a \textit{universal} characterization of topological charges in quantum gravity, further motivating the use of bordism in characterizing the ``charges'' of objects whose tensions we wish to constrain. 

\subsection{Review of Cobordism Conjecture} \label{ssec:review-cobordism-conjecture}

In this section, we briefly review the CC \cite{mcnamara_cobordism_2019} with a focus on the general features of topological charges in quantum gravity. The Cobordism Conjecture (CC) states that any two configurations in quantum gravity (with the same non-compact spacetime asymptotics) are connected to one another with a finite action process. More precisely, suppose we are given any two $d$-manifolds $M$ and $M'$ equipped with the appropriate quantum fields $\C{F}$ for a consistent theory of quantum gravity.
Let $W$ be a $(d+1)$-manifold, equipped with the same quantum fields $\C{F}$ such that $\p W = M + \bar{M'}$ (where $\bar{M'}$ is the orientation reversal of $M'$) in such a way that the fields on $W$ are compatible with the boundary conditions over $M, M'$. We call such a manifold $W$ a \textit{bordism} between $M, M'$. 

Given a tangential structure $\chi$\footnote{By ``tangential structure'', we mean whatever fields/conditions are needed to assume about the manifolds for it to make sense for the particular theory at hand. This can involve more data than just the literal tangential structure, but we still use this abbreviation when referring to what is needed for QG to make sense.}, we define the \textit{bordism group} $\Omega_d^\chi$ of closed $d$-manifolds with tangential structure $\chi$ to be the group of bordism equivalence classes of $d$-manifolds $[M]$ with group operation disjoint union, where we say two manifolds $M, M'$ are equivalent iff there exists a bordism $W$ between $M$. The CC for a $D$-dimensional quantum gravity theory is then concisely expressed as follows: 
\begin{equation} \Omega_d^{\m{QG}} = 0, \q 0 \leq d \leq D \end{equation}
where $\m{QG}$ is ``the tangential structure associated to quantum gravity'' for a suitable defintion thereof. The CC is argued from the absence of global symmetries in quantum gravity -- were there to be a nontrivial group $\Omega_d^{\m{QG}}$, one could construct a topological $(D-d-1)$-form global symmetry, the topological charge associated to which is labeled by $\Omega_d^{\m{QG}}$ itself. Thus, by the absence of global symmetries (including $(-1)$-form global symmetries for the $d = D$ case) one expects all of the above bordism groups to vanish in a full theory of quantum gravity.\footnote{It is conjectured furthermore that all \textit{higher} bordism groups vanish as well: $\Omega_d^{\m{QG}}(\m{pt}) = 0$, though there is less direct motivation for this. Possible connections between the vanishing of these higher bordism groups and the absence of $(-p)$-form global symmetries for $p > 1$ have been suggested in \cite{mcnamara_cobordism_2019}, though a detailed study of this remains to be done.} That is, every configuration in quantum gravity is bordant to every other configuration! 

In practice, suppose we are given a $D$-dimensional effective theory of quantum gravity with tangential structure $\chi$. For practically computable tangential structures $\chi$, the bordism groups $\Omega_d^\chi$ rarely vanish. However, this indicates only that a UV-complete theory of quantum gravity must be \textit{supplemented} by additional objects to take care of the global symmetries. Suppose we have a nontrivial bordism group $\Omega_d^\chi \neq 0$ in a consistent EFT for quantum gravity. There are two possible cases: 

\begin{itemize}
    \item The symmetry is \textbf{broken}. In this case, there is a singular bordism trivializing the class of every generator $[M] \in \Omega_d^\chi$. Accordingly, there are dynamical dimension-$(D-d-1)$ singular branes \textit{sourcing} the bordism charge associated to $\Omega_d^\chi$. 
    
    \item The symmetry is \textbf{gauged}. In this case, the group $\Omega_d^\chi$ does not by itself represent a consistent background for the UV quantum gravity completion. We must supplement the EFT theory with additional fields/topological structures via $\chi \to \chi'$ with the effect that in the completed theory, $\Omega_d^{\chi'} = 0$. 
\end{itemize}

We will see examples of both cases later in the text, and each leads to different predictions for the spectrum of defects in the theory. For now, we focus on the case where the symmetry is \textit{broken}, leaving the case of gauged symmetry to Sec. \ref{ssec:gauged-bordism-defects}. In the case of a nonzero bordism group $\Omega_d^\chi$ whose symmetry is broken in the UV-complete theory, we thus predict the existence of several singular $(D-d-1)$-dimensional brane configurations in quantum gravity. We then say that the $(D-d-1)$-dimensional branes are \textit{bordism defects} and/or that they \textit{carry bordism charge} associated to $\Omega_d^\chi$. This terminology derives from the example of magnetic monopoles---A monopole is a trivialization of the bordism class $\Omega_2(BU(1)) = \B{Z}$ of (oriented, spin, etc.) manifolds with a $U(1)$ gauge bundle, generated by the class $[S^2, F_2 = 1]$ of a sphere $S^2$ with one unit of magnetic $F_2$-flux \cite{mcnamara_cobordism_2019}. We say that a monopole carries magnetic charge in that its worldvolume couples to the dual $U(1)$ (higher) gauge field $A_m$ with $\d A_m = *F_2$. This aspect is discussed further in Sec. \ref{ssec:duality-topological-charges}.) It is this perspective which will be the appropriate interpretation of the CC for our purposes, even though it is somewhat imprecise. In particular, in the language of Sec. \ref{ssec:duality-topological-charges}, bordism charges are \textit{detected magnetically}. 

\subsection{Gauged Symmetry and Bordism with Defects} \label{ssec:gauged-bordism-defects}

In this section, we study in further detail the bordism picture of \textit{gauged} symmetries in quantum gravity. Unlike the case of broken symmetry, we no longer necessarily stipulate the existence of defects with bordism charge, since the associated bordism classes are deemed inconsistent backgrounds anyway. However, it will frequently be the case that one can nevertheless consider the trivialization of a background $[M]$ associated to a gauged symmetry by including \textit{relative} defects. We sketch this mechanism in this section; this discussion will becomes extremely relevant in Sec. \ref{ssec:non-susy-heterotic} when we study non-supersymmetric heterotic branes (cf. \cite{kaidi_non-supersymmetric_2024, kaidi_non-supersymmetric_2023}). 

Consider the case of 10d heterotic (either $(E_8 \times E_8) \rtimes \B{Z}_2$ or $\Spin(32)/\B{Z}_2$) string theory in the absence of NS5-branes and gauge bundles. There is a nontrivial Bianchi identity for the $H_3$-flux over a background spacetime $M$ (which for $F=0$ reduces to): 
\begin{equation} \label{eq:dH3-bianchi-identity}
    \d H_3 \propto \tr R \wedge R.  
\end{equation} 
Accordingly, any manifold with nontrivial first Pontryagin class is independently an inconsistent background for the heterotic string altogether. For instance, take the class $[\m{K3}] \in \Omega_4^{\Spin}(\m{pt})$ of K3 in dimension-4 $\Spin$ bordism. We have $p_1(\m{K3}) = -48$, so the global symmetry associated to the bordism group generated by K3 is \textit{gauged} in the full theory, and the K3 background by itself is ruled out. 

However, if we include NS5-brane sources, working out the constant factors in Eq. \eqref{eq:dH3-bianchi-identity} shows that including precisely 24 units of fivebrane charge on K3 is enough to cancel to the contribution from $p_1(\m{K3})$, and this can be described in bordism language as follows. Suppose we excise 24 copies of $S^3$ equipped with a single unit of $H_3$-flux from K3; the resulting \textit{open} manifold is then a consistent background for the theory. We can trivialize the classes $[S^3, H_3 = 1]$ with a \textit{singular} nullbordism around an NS5-brane, producing a (closed) K3 with 24 NS5 singularities sourcing $H_3$-flux. 

This \textit{singular} K3 in the category of bordisms \textit{with defects} is a consistent background for the heterotic string, and the CC predicts that it itself must have trivialization. The resulting defect is a singular dimension-5 non-supersymmetric \textit{relative} defect that is a simultaneous boundary of 24 NS5-branes. A schematic picture of this is shown below in Fig. \ref{fig:gauged-symmetry-bordism}. In fact, this picture where \textit{gauged} bordism symmetries predict relative defects is rather general, and we will see another example of this in Sec. \ref{ssec:non-susy-heterotic} with the ``electric-magnetic dual'' Bianchi identity for the flux $H_7 = * H_3$ in the $\Spin(32)/\B{Z}_2$ heterotic theory. 

\begin{figure}
    \centering
    \includegraphics[width=0.6\linewidth]{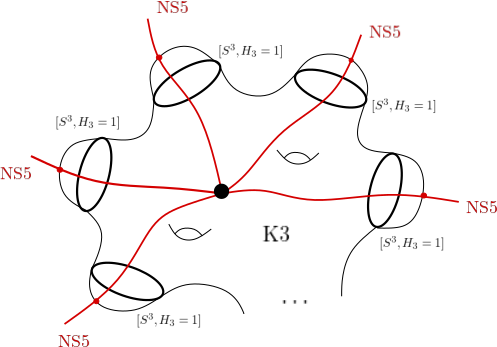}
    \caption{A schematic depiction of a junction defect forming a joint boundary for 24 NS5-branes as a trivialization of a K3 with 24 singularities sourcing $H_3$-flux. The red dots depict the singular nullbordism for $[S^3, H^3 = 1]$ around an NS5-brane, and the black dot represents the fourbrane which corresponds to a singular nullbordism for the entire singular K3 and is relative to the 24 NS5-branes.}
    \label{fig:gauged-symmetry-bordism}
\end{figure}

\subsection{Topological Charges and Bordism} \label{ssec:duality-topological-charges}

In this section we wish to reframe the statement of the CC, thinking instead in terms of the \textit{charges} carried by defects in quantum field theory and quantum gravity. One of the main goals of this paper is to argue from general Swampland and quantum gravitational principles for constraints on the \textit{tension} of an object in terms of its \textit{charge}, but this first requires a precise understanding of what actually constitutes ``charge''. 

In quantum field theory, there are essentially two ways to think about charge. First, one can measure charge \textit{electrically}, which corresponds to viewing the charge as the representation $\rho$ of some gauge symmetry $G$. For instance, the integrality of an electron's charge $n \in \B{Z}$ is demanded by Dirac charge quantization, and should be thought of as a label in $\m{Rep} (U(1)) = \B{Z}$. 

Conversely, one can compute charge \textit{magnetically}. A magnetic object is a submanifold $\Sigma_{d-p-2} \into X$ of codimension-$(p+2)$ in an ambient space $X$ with topologically nontrivial field configurations over the transverse sphere $S^{p+1}$ that label the magnetic charges carried by $\Sigma_{d-p-2}$. Concretely, for a $U(1)$ $p$-form gauge field $A_{p}$ over $S^{p+1}$, the associated Chern class $[F_{p+1}] \in H^{p+1}(S^{p+1}, \B{Z}) = \B{Z}$ labels the magnetic charge carried by $\Sigma_{d-p-2}$. Magnetic charge is naturally generalized to charge \textit{detected by bordism} in the sense that an object carrying magnetic charge may be thought of as a singular trivialization of a bordism class associated to $S^{p+1}$ carrying a single unit of $F_{p+1} \in H^{p+1}(S^{p+1}, \B{Z})$. Accordingly, more general bordism defects can be thought of as ``gravitational'' versions of magnetic charges. We remark that the notions of electric and magnetic charges in quantum field theory are vastly different at first glance, but it is a general feature that gauging an electric $p$-form $G^{(p)}$ symmetry realizes its magnetic $\tilde{G}^{d-p-1}$ dual as a \emph{global} $(d-p-1)$-form symmetry.

It may seem \textit{a priori} like topological charges can be much more general than those detected by bordism, since bordism charges necessarily form abelian groups. However, it appears that quantum gravity washes out apparently nonabelian and noninvertible charges. For example, vortex charges in codimension two need not \textit{a priori} have abelian fusion --- codimension-two submanifolds with gauge holonomy $\Gamma$ along the linking $S^1$ have nonabelian fusion rules (as was discussed extensively in \cite{delgado_finiteness_2024}). However, in \cite{mcnamara_gravitational_2021}, it was argued that nonabelian charges $\Gamma$ carried by codimension-two branes are \textit{reduced} to the abelianization $\m{Ab}(\Gamma) = \Gamma/[\Gamma, \Gamma]$ after including quantum-gravitational effects (where $[\Gamma, \Gamma]$ is the commutator subgroup of $\Gamma$). Briefly, a vortex associated to $aba^{-1} b^{-1} \in [\Gamma, \Gamma]$ dissolves in a ``gravitational soliton'' given by $T^2$ with a holonomy for $a, b$ around each of its 1-cycles and is thus smoothly nullbordant. Similarly, the completeness of electric charge representations for non-abelian gauge groups follows from the fact that there would otherwise be a non-invertible global symmetry \cite{heidenreich_non-invertible_2021}, but as argued in \cite{heckman_fate_2024} these are generically expected to be broken in quantum gravity theories anyway.

\section{Elementary Constituents Conjecture and Motivations} \label{sec:ecc-motivations}

In this section, we move towards a \textit{refinement} of the Cobordism Conjecture (CC). A blind spot of the CC is the fact that it does not speak to the dynamical properties of the bordism defects, and a general dynamical understanding of these objects remains a difficult task, especially in non-supersymmetric settings. However, one can make a few preliminary statements about the \textit{tensions} of such defects. 

First of all, it is possible to create bordism defects with arbitrarily large charge and tension. For instance, take the class of $S^5$ in Type IIB string theory equipped with $N$ units of $F_5$-flux. The associated object carrying this bordism charge is a stack of $N$ D3-branes, which has tension $T \simeq N$ in Planck units. This is consistent with the Weak Gravity Conjecture (WGC), and indeed, the WGC predicts a bound on the tension of objects with arbitrarily large charge, but the price we pay is that the \textit{tension} may also grow arbitrarily large. However, the WGC applies only to $U(1)$ gauge charges which carry bordism charge of the form $[S^k_{F_k}]$, spheres $S^k$ equipped with flux $F_k$. For more general bordism defects, and especially \textit{torsion} charges, the motivation for WGC coming from extremal black holes fails to make sense. 

Given that it is not generally possible to characterize a charge by comparing it to some extremal black hole background in classical gravity, we want an intrinsically \textit{quantum-gravitational} approach to constraining the tensions of bordism defects. Accordingly, the appropriate place to look is at objects with \textit{small} charge; namely, the \textit{generators} of bordism defects. In this case, we expect the associated objects to not be ``too heavy''; if they were too heavy, they could form a black hole with large horizon area, and thus a large microstate count. This would contradict the intuitive idea that bordism \textit{generators} are ``indecomposable'' in that they cannot be broken into smaller chunks. Accordingly, we expect the objects carrying ``unit charge'' that trivialize bordism groups to be fully quantum and without classical horizons. The only appropriate scale against which their tension can be compared is then $m_{\m{pl}}$, and we are thus motivated to think of bordism generator defects as being (sub-)Planckian objects.

In the remainder of this section, we formalize this idea. In Sec. \ref{ssec:conj}, we formally introduce our conjecture (Conjecture \ref{conj:ecc}), which we call the Elementary Constituents Conjecture (ECC). In Secs. \ref{ssec:examples-m-theory} and \ref{ssec:examples-d-branes}, we study simple examples that confirm our intuition for Conjecture \ref{conj:ecc}. These examples, particularly in Sec. \ref{ssec:examples-d-branes}, illustrate the necessity for an \textit{attractor mechanism} sending the moduli to an appropriate regime where the tension of a bordism generator is (sub-)Planckian. 

\subsection{The Conjecture} \label{ssec:conj}

In this section, we move towards a precise formulation of a conjecture which constrains the tensions of bordism generators. From the previous discussion, we have the intuitive picture, but an immediate issue presents itself. The major subtlety is the idea that tensions in quantum gravity are sensitive to continuous deformations of the theory; that is, they are \textit{moduli-dependent}. We argued previously that the tension of bordism generators should be (sub-)Planckian. However, it only truly makes sense to say that this should be true \textit{in some regime of the moduli space}. 

So how do we ``find'' the region where the tensions of bordism generators are indeed (sub-)Planckian? The answer rests in a (generalization of a) key feature of the study of supergravity solutions of extremal black holes --- the \textit{attractor mechanism}. The scalars in the full supergravity solution in the presence of the defect are not constant at their asymptotic values; they \textit{flow} towards universal fixed points near the core of the defect. These fixed points turn out to point in exactly the directions in moduli which \textit{decrease} the tension of the object. Thus, even if we start at a value of asymptotic moduli where a bordism generator is \textit{heavy} relative to Planck, the attractor flow pulls the moduli in the right direction. 

In summary, we may formalize our more sophisticated understanding of the situation in our following conjecture: 

\begin{conj}[Elementary Constituents Conjecture] \label{conj:ecc}
For every bordism group, there are generator defects that trivialize the group with the following (interrelated) properties:
\begin{itemize}
    \item The generators are \emph{elementary} in that their (suitably defined) entropy density 
    goes as $s \simeq 1$. This entropy density can also be viewed as the measure of the complexity of the theory living on the defect generator (such as a $c$-function). Correspondingly, bordism generators do not have (classically describable) horizons.
    \item There is a low energy EFT description of the theory (with an appropriate UV cutoff) such that there is a region in the EFT field space in which defect generators have tension $0\leq T \leq  O(1)$ in Planck units. \footnote{The lower bound is not trivial. In some examples (cf. Sec. \ref{ssec:orbifold-orientifold}), non-torsion bordism classes can be represented by objects with negative tension. However, even in these cases there exist generators with nonnegative tension. Our conjecture posits the existence of such objects in \textit{all} cases.} 
    \item As one approaches a defect generator, a (suitably generalized version of the) attractor mechanism drives all of the fields in the theory towards the region of field space leading to $0\leq T \leq O(1)$.
\end{itemize}
\end{conj}

\subsection{Examples in M-theory} \label{ssec:examples-m-theory}

We now move to some examples which motivate our conjectures. Let us first look at the simplest possible case -- BPS objects in M-theory on oriented manifolds. The most obvious objects are the M2- and M5-brane, which couple to the electric and magnetic charge associated to the $C_3$ gauge field. It may be clear that these are solitonic bordism generators, but let us take a moment to clarify precisely which bordism charges they carry. For M-theoretic backgrounds, we have Dirac-quantized $G_7$- and $G_4$-flux transverse to M2- and M5-branes, respectively. In this language, the M2- and M5-branes carry charge associated to the integer-valued bordism classes 
\begin{align} [S^7_{G_7}] & \in \Omega_7(K(\B{Z}, 7)) \\ 
[S^4_{G_4}] & \in  \Omega_4(K(\B{Z}, 4))
\end{align}
of $S^7$ and $S^4$ equipped with one unit of $G_7$- and $G_4$-flux, respectively, which in turn generate (factors of the) respective bordism groups.\footnote{Here we adopt the ``naive'' Dirac quantization for the M-theoretic fluxes $G_4$ and $G_7$, which ignores the effect of the Chern-Simons term $C_3 \wedge G_4 \wedge G_4$ as well as the gravitational term $C_3 \wedge I_8(R)$. We return to subtler topological characterizations of M2- and M5-brane charge in Sec. \ref{sssec:m-theory-orientifolds}, and for a full treatment, it is worth mentioning a proposed unifying model for M-theoretic flux quantization in (unstable) twisted cohomotopy \cite{sati_flux_2025}.} (Recall that $K(\mathbb{Z},n)$ is the classifying space for $(n-1)$-circle bundles with higher-form gauge fields $C_{d-n-1}$ in a $d$-dimensional theory. For example, $\pi_7(K(\B{Z}, 7))$ should be understood as the higher gauge field configuration sourced by the M2-brane on the $S^{7}$ it links.) 

There are no moduli in M-theory, and the M2- and M5- tensions are therefore constant in Planck units. Even here, however, it must still in principle be checked what the tension is in Planck units, since we propose a bound $T \lesssim 1$. Though we do not specify what the exact ``order 1'' factor is, it is instructive to calculate what these factors are in this simple example. For the M-theory effective action, we have
\begin{equation} S_{11} = \frac{1}{2\kappa^2}  \int d^{11}x \, \sqrt{-G} \left( R - \frac{1}{2} |G_{4}|^{2} \right)
- \frac{1}{6} \int C_{3} \wedge G_{4} \wedge G_{4} + \cdots, \end{equation}
where
\begin{equation} 2\k^2 = \frac{1}{2\pi}(2\pi l_{\m{pl}, 11})^9. \end{equation}
We find also that 
\begin{equation} T_{\m{M2}} = 2\pi(2\pi l_{\m{pl}, 11})^{-3}, \q T_{\m{M5}} = 2\pi(2\pi l_{\m{pl}, 11})^{-6}, \end{equation}
and in units where $\k^2 = 1$, 
\begin{equation} T_{\m{M2}} = \frac{1}{2}(4\pi)^{2/3}, \q T_{\m{M5}} = \frac{1}{2} (4\pi)^{1/3}. \end{equation}
These constants are appropriately $O(1)$ in that they are not large enough to back-react and induce a classical gravitational background. Moreover, since we are considering single M2- and M5-branes as generators of bordism charge, their worldvolume theories will have entropy density $s \simeq 1$ (coming, for instance, from the appropriate horizon area formula of the $\m{AdS}_4 \times S^7$ and $\m{AdS}_7 \times S^4$ geometries). In all subsequent examples, we will not calculate these explicit factors, but we recognize that they will be similar to the M2- and M5-brane tensions computed here.

\subsection{Examples with Type II D-branes} \label{ssec:examples-d-branes}

We now turn to the next simplest example of objects carrying bordism charge --- D-branes. (We focus in this section on D-branes, but we remark that other BPS solitons with a spherical horizon, such as the NS5-brane, may be treated entirely analogously.) In Type II string theory, (supersymmetric) D-branes carry RR charge. In more detail, a D$p$-brane is transverse to a sphere $S^{8-p}$ carrying (integer-quantized) RR flux $F_{8-p}$. The associated bordism charge is given (for instance) by 
\begin{equation} [S_{F_{8-p}}^{8-p}] \in \Omega_{8-p}^{\Spin}(K(\B{Z}, 8-p)), \end{equation}
generating an integer-valued bordism group.\footnote{Here, we forgo the more proper K-theoretic characterization of D-branes in favor of a simple model using classes in ordinary integer cohomology. We return to K-theory in Sec. \ref{sssec:type-i-dbranes}, where it becomes essential to properly characterize non-supersymmetric D-branes in Type I.} 
However, D-branes are different from the M2- and M5-brane discussed previously in one crucial respect --- their tensions are now dependent on the dilaton modulus, given by $g_s = e^\Phi$. Accordingly, in an appropriate regime of $g_s$, the tension of a D-brane becomes very large, and the object classically backreacts on the geometry. The corresponding solutions are supergravity \textit{black branes}, which feature a running profile for the dilaton as a function of the radial distance to the brane.

In this context, we encounter so-called ``dynamical cobordism'' \cite{buratti_dynamical_2021-1} or ``running'' solutions, where as we approach the near horizon limit, the value of scalar moduli runs off to large distances in the corresponding moduli space. For D-branes, one finds the \textit{string} frame tension, $T \sim 1/g_s$, and this scaling can be expressed in the Einstein frame in Planck units by relating $m_{pl,10}$ and $\alpha'$ via $g_{s}$. We obtain the following expression for the tension of a D$p$-brane in Planck units that now depends on the dimension $p+1$ of the worldvolume because of the normalization of the Dirac-Born-Infeld (DBI) action:
\begin{equation} T_p = g_s^{(p-3)/4} . \end{equation}
Already, we conclude that for $p < 3$, the D-branes become light relative to Planck as $g_s \to \infty$, while for $p > 3$, the D-branes are light as $g_s \to 0$. For $p = 3$, D-brane tension is fixed at $T = 1$, analogous to the M2- and M5-brane cases. All of these cases are clearly consistent with Conjecture \ref{conj:ecc}. As one approaches the brane, one finds the following behavior: 
\begin{itemize}
    \item For $p > 3$, we have $\Phi \to -\infty$ as $r \to 0$, so that $g_s(r) \to 0$ and also $T_p(g_s) \to 0$. 
    \item For $p < 3$, we have $\Phi \to + \infty$ as $r \to 0$, so that $g_s(r) \to \infty$ but also $T_p(g_s) \to 0$.
    \item For $p = 3$, $T_p$ remains fixed at $m_{\m{pl},10}^4$ and independent of the value of the dilaton. 
\end{itemize}
In all cases, we therefore observe that the ``core tension'' of the requisite object becomes sub-Planckian as one approaches $r \to 0$, and the $p=3$ case is special precisely because the $D3$-brane is self-dual under the S-duality of Type IIB, and its near horizon geometry backreacts to yield an $\text{AdS}_{5} \times S^{5}$ geometry. 

We emphasize that the ``moduli-dependent tensions'' $T_p(e^\Phi)$ are \textit{not} physical tensions in a literal sense. Instead, we merely conclude that the moduli are driven in a direction that, were the asymptotic moduli taken to be the attractor values, the tension of the brane would be sub-Planckian (and indeed, approaching zero). We study this setup in further detail in Appendix \ref{app:stretched-horizon-entropy}, clarifying also the subtleties associated with approaching within Planck distance of the brane. 

There is a large class of examples -- supersymmetric and non-supersymmetric -- corresponding to D-branes wrapping cycles in the internal geometry of a string compactification. The corresponding objects in the lower dimensional theory will have their tension or mass controlled by the (Einstein frame) volume of the corresponding cycles, the generators of which we claim are sub-Planckian at the attractor value. For example, suppose we compactify M-theory on a Calabi-Yau 3-fold (CY3) to obtain a 5d $\mathcal{N}=1$ theory. There is a region in the Kähler cone where BPS black holes exist, and as we approach the boundary of this region, the corresponding black holes have near-zero horizon area and are very light. \footnote{Even if some of the generators on the boundary of this region are not BPS, their mass is not expected to be much different from the BPS bound, which is all that the ECC requires. Note also that if the Kähler class is one dimensional, the BPS lattice is also one dimensional, and the minimum mass in Einstein frame is of order of Planck. This illustrates that typically we either have a Planck mass generator (with a single modulus) or nearly massless generators (in cases with more Kähler moduli).} This is already an example in which a generating set of a class of states is (nearly) massless. More generally, in the presence of multiple volume moduli, it is expected that a given cycle can be shrunk relative to the total volume of the internal manifold, furnishing a regime where the tensions of associated defects are sub-Planckian. An interesting class of exceptions is explored in Sec. \ref{ssec:implications-geometry}, where volume moduli are absent and the ECC thus makes a prediction for the volumes of such ``rigid'' cycles.

\section{Attractor Mechanism} \label{sec:attractor}

In this section, we study in more detail the nature of  attractor flows in scalar field spaces $\mathcal{M}$. In doing so, we highlight the interpretation of attractor fixed points as minimizing some quantity associated to an object. This quantity, in various examples, is to be understood as either the \textit{tension} or \textit{entropy} associated to the singular object. 

The principal idea behind the attractor mechanism is the \textit{separation} of the tension of a singular solution to the EFT equations of motion into two pieces: An ``intrinsic'' tension at the core of the object and some ``extrinsic'' field configuration outside the object carrying some additional energy. The attractor mechanism, at least for BPS objects, yields a profile for the scalar fields in the theory in such a way that the ``intrinsic'' \textit{core tension} of the object is minimized. That is, the scalar fields approach a value $\Phi_0$ as one approaches the core of the brane such that, were the asymptotic values of the scalars equal to $\Phi_0$, the \textit{asymptotic tension} as measured by an observer at infinity would be minimal. 

\subsection{Review of BPS Attractor Mechanism}

As a preliminary,  we briefly review the idea of BPS attractor flows as seen in string theory \cite{ferrara_n2_1995,moore_attractors_2003, ferrara_black_1997,  behrndt_classical_1997,ferrara_supersymmetry_1996,ruehle_attractors_2024}. Typically, the central charge $Z$ of a BPS object depends both on a charge vector, obtained from a BPS lattice, and various scalar moduli parameterizing internal volumes and couplings in a certain string background. The presence of a static BPS object induces a running solution for the various effective fields in theory, and the BPS attractor flow fixes the value of the running moduli at the location of the object in terms of only the BPS charges. This value of moduli may or may not lie at finite distance in moduli space: In the case of black $p$-branes, for example, the dilaton runs off to the boundary of moduli space for $p \neq 3$.

To give a very concrete example, consider a BPS black hole in a $4d$ $\mathcal{N}=2$ supergravity theory, which can be obtained by wrapping D$p$-branes on a calibrated $p$-cycle in a Calabi-Yau compactification of Type II string theory. The static, spherically symmetric black hole solution in 4d is
\begin{equation} \label{sq:spherical-metric} ds^2 = -e^{2U} \d t^2 + e^{-2U} (\d x^i)^2, \end{equation}
where the $x^i$ denote transverse directions and $U = U(r)$ is a function of the radial distance $r = \abs{\vec{x}}$ alone. It is conventional to define the auxiliary radial coordinate $\tau = r^{-1}$, so that spatial infinity lies at $\tau = 0$. The BPS conditions are expressed as first-order equations in the variables $(U, \phi^a)$, where $\phi^a$ are local coordinates on the moduli space of scalar fields in the supergravity effective action:
\begin{align}
    \frac{dU}{d\tau} &= -e^U \abs{Z} \\
    \frac{d\phi^a}{d\tau} &= -2 e^U g^{a\bar{b}} D_{\bar b} \abs{\bar Z}.
\end{align}
Here $g_{a\bar b}$ is the Weil-Petersson metric on the scalar moduli space $\C{M}$, appearing as the coefficient of the scalar kinetic term $\frac{1}{2} g^{ab} (\p_\u \phi^a)(\p^\u {\bar\phi}^{b})$ in the effective action, and $Z$ is the BPS central charge, expressed as a function of the moduli $\phi^a$. The flow equations for $\phi^{a}$ above are gradient flows set by the central charge $Z$, and drive the latter to a minimum at the horizon of the black hole. 

More generally, one can interpret the BPS attractor flow in two equivalent ways. From one perspective, it is merely the statement that a BPS object is invariant under the supersymmetry it is supposed to preserve, implying the condition $D_{i} \abs{Z(\Phi_{0}, \Gamma)}  = 0$. Here $D$ is the is covariant derivative with respect to the Kahler connection on the moduli space $\mathcal{M}$, and  $\Phi_{0} \in \mathcal{M}$ is the moduli at the horizon.

From another perspective, the BPS condition establishes a relation between tension and central charge
\begin{equation}\tau = \abs{Z(\Phi_{\infty})}, \end{equation}
where $\Phi_{\infty} \in \mathcal{M}$ is the collection of asymptotic moduli in the supergravity effective action. The tension so defined is analogous to an ADM mass, as necessitated by general covariance. This means that the energy density associated to the profile of various fields sourced by the BPS object will contribute to this quantity.  The attractor fixed point $\Phi_{0} \in \mathcal{M}$ is such that the quantity $\abs{Z(\Phi_{0})}$ is minimized for a fixed set of BPS charges. The interpretation of this minimization is that a maximum amount of ADM tension $\tau$ should be acquired from the energy density of fields outside the core of the object, where this maximization is performed with the BPS charges as a constraint. In other words, the attractor fixed point gives the core tension of the BPS object, and it makes sense that this quantity only depends on charges and not on asymptotic moduli. 

It is this latter interpretation that we will adopt, and we emphasize that the attractor flow minimizes a quantity that is intrinsic to the object being studied. This idea of minimization, involving tensions and constraints given by charges, seems very close in spirit with the action principle itself, and motivates a more general phrasing of attractor mechanism that depends on equations of motion in the vicinity of an object, to which we now turn. 
 
\subsection{Generalized Attractor Mechanism}

In \cite{sen_black_2008}, Sen proposed a more general, not necessarily supersymmetric, notion of an attractor mechanism for extremal spherical black holes in $D$ spacetime dimensions. The mechanism proposed consists of extremizing an effective action in the near horizon geometry of the black hole, subject to appropriate constraints. It turns out that this formalism only truly depends on the isometries of the near-horizon metric being of the form $SO(2,1) \times SO(D-1)$, in which the enhancement from $SO(1,1)$ to $SO(2,1)$ in the near-horizon isometry group is due to extremality. The use of this isometry is to decompose the metric $g_{\mu \nu}$ and other fields into pieces whose coefficients can then be obtained by extremizing a certain quantity. 

The quantity to be extremized is the 2d effective action obtained by integration of the $D$-dimensional Lagrangian $\C{L}$ over the radial and time directions: 
\begin{equation} I = \int \d t \d (r^{-1}) \sqrt{-\det h} \C{L}, \end{equation}
where $h$ is the spacetime metric reduced over the angular directions. 

In this case, the action above is a functional of the scalars parameterizing the metric along with the various gauge fields sourced by the black hole (or brane, more generally). We define a function $f(\Phi, \Gamma)$ by
\begin{equation} f(\Phi, \Gamma) = I,  \end{equation}
where $\Phi$ is the collection of scalar moduli fields of which $\C{L}$ is a functional and $\Gamma$ encodes charges. In the case discussed by Sen, $\Phi = (\vec{v},u)$   where $\vec{v}$ parameterize the metric, and $u$ the additional scalar fields. The minimization problem he considers also specifies $\Gamma =  ( \vec{e},\vec{p})$, in which  $\vec{e}$  parameterizes the gauge field strengths in the $r-t$ plane, and  $\vec{p}$ the field strengths on the angular linking sphere. In that case, $\Gamma$ is the charge vector associated to the electric and magnetic charges of the spherically symmetric extremal black hole.  

For eternal solutions, there is no kinetic term and the extremization reads
\begin{equation} \dfrac{\partial f}{\partial \Phi} = 0, \end{equation}
which exactly fixes the metric and (some) scalar moduli. The parameters labeling gauge fields enter through $\Gamma$, and further extremizing the gauge sector will give the known electric and magnetic charges of the extremal objects. In this picture, the extremization enacted by the attractor mechanism minimizes the \emph{tension} associated to the core of the object.

When talking about elementary constituents, one may also want to consider entropy, as discussed in Sec. \ref{sec:ecc-motivations}. Accordingly, there is an entropy-like quantity that can be defined directly from this generalized attractor flow

\begin{equation} \mathcal{S}(\Phi, \Gamma, f) = p_{i} \dfrac{\partial f}{\partial e_{i}} - f(\Phi, \Gamma) ,   \end{equation}
which is a Legendre transform of $f$, with $\p f / \p e_{i} = q_{i}$ being the electric charge. This quantity is nothing but the Hawking-Bekenstein entropy of the corresponding extremal horizon, and the attractor flow can be understood as minimizing this entropy (which does not depend on any asymptotic moduli by nature of the Legendre transform). Even though in this section we have focused on generalized attractor mechanism for extremal black holes, similar story repeats itself for extremal black branes.

\section{Supersymmetric Examples} \label{sec:susy-examples}

In this section, we review various further examples that offer several new perspectives on our conjecture. In addition to the elementary examples with D-branes and M-branes, we find a much wider variety of defects carrying bordism charge whose tensions can be analyzed and understood, and many of these objects have \textit{a priori} surprising properties that align with our conjecture in nontrivial ways. 

We first investigate examples of the ECC for objects of codimension two in Sec. \ref{ssec:codimension-leq-2}. The treatment of codimension-two branes requires some care, as classical solutions thereof typically have large backreaction and no well-defined asymptotic geometry. For example, the supergravity solution for a D7-brane in flat spacetime naively appears to carry \textit{infinite} tension due to the energy stored in the varying axiodilaton profile. However, this is remedied by recognizing that a D7 is not \textit{independently} a consistent background in flat space at all --- instead, one must study the fully backreacted uplift to F-theory \cite{greene_stringy_1990}, where the D7-brane tension is identified with the deficit angle it sources on a compact background. As per the ECC, we find that we can always choose generators with nonnegative deficit angles (of order 1). 

We next study (supersymmetric) \textit{orientifold} examples in Sec. \ref{ssec:orbifold-orientifold}, which have been understood as bordism generators of unoriented manifolds \cite{montero_cobordism_2021}. A curious feature of orientifolds is their apparent \textit{negative} tension. However, as for the \textit{bordism} charge carried by orientifolds, we find that we can always choose nonnegative-tension generators of the appropriate bordism groups, in agreement with the ECC. We comment further on non-perturbative generalizations of orientifolds in F-theory -- \textit{(singular) Non-Higgsable Clusters} and \textit{S-folds} and argue also for the existence of nonnegative-tension sub-Planckian bordism generators. In these examples, we call the chosen nonnegative-tension generators \textit{ECC-consistent}. 

In addition to the main examples in this section, we also revisit moduli-dependent tensions in a much more general setting in Appendix \ref{ssec:compactifications-maximal-susy}. By studying the ``purely saxionic'' slice in the moduli space of M-theory on a rectangular $T^k$, we revisit the arguments in \cite{obers_u-duality_1999, iqbal_mysterious_2001} and argue that there is a general interior region of the moduli space around the ``self-U-duality'' point at which the tensions of the \textit{entire} $\frac{1}{2}$-BPS spectrum are Planckian, further motivating the ECC. 

\subsection{Examples in Codimension Two} \label{ssec:codimension-leq-2}

In this section, we study the tensions of various branes in codimension two in string theory, with a focus on sevenbranes in Type IIB string theory. The \textit{tension} of a codimension-2 defect is rather special in that it can be deduced exactly from a singular classical solution: In Planck units, it is nothing other than the deficit angle of the solution. Explicitly, suppose a codimension-2 vortex defect is localized at the origin, corresponding to the following conical singularity: 
\begin{equation} ds^2 = \d r^2 + r^2 d\theta^2, \q \theta \sim \theta + \b, \end{equation}
for some angular periodicity $\b < 2\pi$. Then the tension is given by
\begin{equation} T \simeq \delta \cdot m_{\m{pl}, d}^{d-2}, \end{equation}
where $\delta = 2\pi - \b$ is the conical deficit angle. 

The description of vortex tension in terms of the associated deficit angle already makes it clear what the criterion for compatibility with the ECC should look like for vortices: We wish for the deficit angle to satisfy $0 \leq \delta < 4\pi$. For \textit{strictly positive} deficit angles, we conclude that only finitely many such defects may be consistently placed on a background before it ``closes in'' on itself as a sphere. 

In more detail, recall the Gauss-Bonnet theorem, which states that for a closed surface $M$ of genus $g \geq 0$ equipped with conical singularities with deficit angles $\delta_i$, we have
\begin{equation} \int_{M} R + \sum_i \delta_i = 2\pi (2 - 2g), \end{equation}
Only for $g = 0$, therefore, is it possible to obtain a positive value on the right hand side. For an ECC-consistent bordism generator with $\delta > 0$, we therefore expect that we can only consistently envision a spherical background with $N$ conical singularities, such that $N \delta = 4\pi$. For ECC-consistent generators with zero deficit angle, one can envision a nullbordism as a surface of $g = 1$ or $g = 0$ (depending on additional contributions from the varying moduli fields), but there is no restriction on torsionality. 

Let us now study explicitly study these phenomena in the example of sevenbranes in Type IIB string theory, investigating their bordism charges and tensions in detail. In Type IIB, the scalar axiodilaton modulus $\tau = C_0 + ie^{-\phi}$ admits an action by $\SL(2, \B{Z})$ via 
\begin{equation} \tau \longmapsto \frac{a\tau + b}{c\tau + d}, \q \mt{a & b \\ c & d} \in \SL(2, \B{Z}) \end{equation}
which is expected to be a gauge symmetry of the full UV-complete theory. The $\SL(2, \B{Z})$ duality acts on the charged spectrum of $(p, q)$-strings in the theory via the action on their electric couplings $B_2, C_2$
\begin{equation} \mt{B_2 \\ C_2} \longmapsto \mt{a & b \\ c & d } \mt{B_2 \\ C_2} . \end{equation}
Accordingly, associated to every consistent background for Type IIB, one expects the data of a $\SL(2, \B{Z})$ \textit{duality bundle} over spacetime. 

This, however, is not the full story \cite{debray_chronicles_2023, tachikawa_why_2019}, as UV consistency with Type IIB string theory reqauires that the $\SL(2, \B{Z})$ duality bundle admit a lift corresponding to the action of the Type IIB duality group on fermions. The tangential structure of a manifold admitting a consistent $\Spin(n)$-lift of the duality group $\SL(2, \B{Z})$ corresponds to a lift of the frame bundle to
\begin{equation} G = \frac{\Spin(n) \times \m{Mp}(2, \B{Z})}{\B{Z}_2}, \end{equation}
where the $\B{Z}_2$ is the diagonal central $\B{Z}_2$ in $\Spin(n)$ and the $\B{Z}_2$-central extension $\m{Mp}(2, \B{Z})$ of $\SL(2, \B{Z})$. As in \cite{debray_chronicles_2023}, we denote the bordism groups with this structure by $\Omega_k^{\Spin\m{-Mp}(2, \B{Z})}(\m{pt})$. Focusing again on $k= 1$, we compute
\begin{equation} \Omega_1^{\Spin\text{-Mp}(2, \B{Z})}(\m{pt}) = \B{Z}_{24}, \end{equation}
where the single generator of $\B{Z}_{24}$ is given by $[S_p^1(\hat{T})]$, a $S^1$ with periodic spin structure and compatible monodromy by the lift $\hat{T} \in \m{Mp}(2, \B{Z})$ of $T \in \SL(2, \B{Z})$. 

Let us now study the generator associated to $[S^1(\hat{T})]$. This nothing but the Type IIB D7-brane, which sources a profile for $\tau$ in such a way that $\tau \mapsto \tau + 1$ as one winds around the D7-brane. The tension of the D7-brane, as we argued, comes from its deficit angle. However, an ADM-style computation of the tension also yields contribution from the background axiodilaton gradient. For a general seven-brane effecting a monodromy by the matrix
\begin{equation} e^Q = \exp\left( \mt{\frac{r}{2} & p \\ -q & -\frac{r}{2} } \right) \in \SL(2, \B{Z}), \q p, q, r \in \B{R} \end{equation}
we compute \cite{bergshoeff_iib_2007} that
\begin{equation} \label{eq:sevenbrane-tension} T_{Q} =  2\delta + \int \frac{i}{2} dz \wedge d\bar{z} \frac{\p \tau \bar\p \bar\tau}{(\m{Im} \, \tau)^2},   \end{equation}
where $\delta$ is the deficit angle of the seven-brane solution, given by 
\begin{equation} \label{eq:deficit-angle} \delta = \m{sign}(q) \sqrt{\det Q}, \q \det Q > 0. \end{equation}

For the D7-brane, the axiodilaton gradient contribution to the tension in Eq. \eqref{eq:sevenbrane-tension} has an infrared divergence. This is due to the divergent volume of the strip $\B{H}/\B{Z}$ obtained by quotienting the upper-half-plane $\B{H}$ by the action of $\tau \mapsto \tau + 1$. (In the language of \cite{delgado_finiteness_2024}, this region is not \textit{compactifiable}.) Recall, however, that our discussion of the attractor mechanism makes it clear that we should separate the tension into a ``core'' contribution (given by the deficit angle $\delta$) and a contribution from the scalar field profile. The formula Eq. \eqref{eq:sevenbrane-tension} for the sevenbrane tension already makes this splitting manifest. For a D7-brane, we have $q = r = 0$, yielding $\det Q = 0$ and a core tension of zero:
\begin{equation} T^{\m{core}}_{\m{D}7} = 2\delta_{\m{D}7} = 0. \end{equation}
Note, however, that (unlike the D$p$-brane examples for $p < 7$) we cannot literally speak of this ``core tension'' as any physical D7-brane tension, since it is impossible to construct a background in which the axiodilaton gradient is absent.  

To summarize, we have seen that a single D7-brane carries zero ``core tension'' but is necessarily accompanied by a varying profile for $\tau$ which yields formally divergent tension on its own. At first, the zero core tension seems to imply that the D7-brane may carry an integer-valued bordism charge. This is further supported by the associated RR-flux carried by $F_1$ in ordinary integer cohomology, which lifts to an (integer-valued) $\KU$-theoretic class. However, in the fully non-perturbative theory, the integrality of D7 charge is illusory, as we have seen that the D7-brane is \textit{torsion} in the bordism of manifolds with a full $\SL(2, \B{Z})$ duality bundle, carrying charge labeled by $\B{Z}_{24} = \Omega_1^{\Spin\text{-}\m{Mp}(2, \B{Z})}(\m{pt})$. 

This torsionality is explained by the contribution of the scalar field profile to the overall D7 tension on a compact background. In more detail, for a configuration of $n$ (not mutually local) D7-branes with a scalar field profile for $\tau$ that presents an $n$-fold covering of the fundamental domain $\B{H}/\SL(2, \B{Z})$, the total tension goes as
\begin{equation} T \simeq n \cdot \m{area}(\B{H}/\SL(2, \B{Z})) + 2 \sum_i \delta_{\m{D7}} = n \cdot \m{area}(\B{H}/\SL(2, \B{Z})) .  \end{equation}
We see therefore that a configuration of sevenbranes acquires finite tension from the axiodilaton profile, proportional to the quantity $\m{area}(\B{H}/\SL(2, \B{Z})) = \frac{\pi}{3}$. The finiteness of the tension of such a configuration, which itself carries bordism charge, derives from the finiteness of the moduli space volume as argued in \cite{delgado_finiteness_2024}.\footnote{In \cite{delgado_finiteness_2024}, an argument was presented for the weaker condition of \textit{compactifiability}, but all known examples of infinite-volume compactifiable moduli spaces have no nontrivial dualities with associated vortices anyway.} The associated ``effective deficit angle'' of a configuration of $n$ D7-branes is then given by 
\begin{equation} \delta_{\m{eff}} = \frac{n}{2} \cdot 
\m{area}(\B{H}/\SL(2, \B{Z})) = \frac{\pi n}{6}. \end{equation}
For $n = 24$, we see that $\delta_{\m{eff}} = 4\pi$, corresponding to a copy of $\B{P}^1$ equipped with a 24-fold covering of $\B{H}/\SL(2, \B{Z})$ and therefore 24 D7-branes; this presents the required nullbordism for $24 [S^1(\hat{T})]$. 

It should be noted that bordism generators associated to duality vortices need not be torsion-valued at all --- there are examples in 4d $\C{N}=2$ theories where the vector-multiplet moduli space satisfies $\C{M} \simeq \B{P}^1 - \{0, 1, \infty\}$ (with compactification $\bar{\C{M}} = \B{P}^1$) and has associated duality group $\Gamma = \pi_1(\C{M}) = \B{Z} * \B{Z}$, for which $\Omega_1^{\Spin}(B\Gamma) = \m{Ab}(\Gamma) = \B{Z} \oplus \B{Z}$ \cite{delgado_finiteness_2024, brav_thin_2014}. In such cases, an analogous analysis shows that the corresponding duality vortices have zero core tension (just like the D7-brane core tension), as they must.

Finally, recall that there is an additional extension of the duality structure of Type IIB due to an extension of $\SL(2, \B{Z})$ by the outer automorphism $\B{Z}_2$. Physically, this corresponds to the action of the worldsheet parity operator $\Omega$, and the combined $\Spin$-lift and outer-extension of $\SL(2, \B{Z})$ is denoted in \cite{debray_chronicles_2023} by $\GL^+(2, \B{Z})$. Again, the full tangential structure corresponds to $(\Spin(n) \times \GL^+(2, \B{Z}))/\B{Z}_2$, and the corresponding bordism group is
\begin{equation} \Omega_1^{\Spin\text{-}\GL^+(2, \B{Z})}(\m{pt}) = \B{Z}_2 \oplus \B{Z}_2. \end{equation}
The associated generators are $S^1(\hat{T})$ and $S^1(\hat{R})$, where $\hat{T}$ is the ($\Spin$-lift of the) generator of the axionic shift $T : \tau \mapsto \tau + 1$ and $\hat{R}$ is the ($\Pin^+$-lift of the) new generator of the $\B{Z}_2$-extension of $\SL(2, \B{Z})$ to $\GL(2, \B{Z})$.\footnote{In \cite{debray_chronicles_2023}, the BPS generator of the first $\B{Z}_2$ factor is described as $[S^1(\hat{S})]$ instead of $[S^1(\hat{T})]$. However, these two generators are actually equivalent in bordism, since the generator $\hat{U} = \hat{S} \hat{T}$ is trivialized by a ``gravitational soliton''. See \cite{mcnamara_gravitational_2021} for details.}
The usual BPS seven-brane sources the usual bordism charge associated to $[S^1(\hat{T})]$, and has zero core tension as argued earlier. The non-supersymmetric \textit{R7-brane} sources the charge detected by $[S^1(\hat{R})]$. The R7 is a nonsupersymmetric torsion generator, and the ECC therefore amounts to a (very rudimentary) constraint on the R7-tension/deficit angle. 

Until recently, the R7 has been a rather mysterious object. It has recently been argued \cite{heckman_gso_2025}, however, that the R7 represents a dynamically $S^1$-reduced domain wall between Type IIA and Type IIB string theory, attached to which is a monodromy cut implementing the $\B{Z}_2$-action of the R7 (corresponding to worldsheet reflection $\Omega$ or left-moving worldsheet fermion parity $(-1)^{\m{F}_L}$). Our conjecture therefore may also be rephrased in terms of the statement that this domain wall between IIA and IIB, also predicted by the Cobordism Conjecture, has sub-Planckian tension.\footnote{Further evidence for this claim arises from the observation that the change in GSO projection as one crosses the IIA/IIB domain wall leaves the NSNS degrees of freedom corresponding to the metric unchanged, so at the very least, the IIA/IIB wall does not couple directly to the bulk metric at leading order in $\a'$ in its worldvolume action.} The dynamics of the R7 are still poorly understood, but here we predict at least a rudimentary constraint on its worldvolume tension and entropy.

\subsection{Examples with Orientifolds} \label{ssec:orbifold-orientifold}

In this section, we will discuss the charges and tensions associated to objects which trivialize bordism classes in \textit{unoriented} bordism and generalizations thereof. The structure of the unoriented bordism ring is completely understood:
\begin{equation} \Omega_{\bullet}^{\m{O}}(\m{pt}) = \B{Z}_2[S]\end{equation}
where 
\begin{equation} \label{eq:unoriented-bordism} S \supset \{ [\B{RP}^i] \mid i = 2k, \q k \in \B{Z} \} \end{equation}
(and contains other generators in odd dimensions $i \neq 2^k-1$). Accordingly, the generators of unoriented bordism include products of real projective spaces $\B{RP}^i$, and the associated objects which source topological charge detected by $\B{RP}^i$ are realized as \textit{orientifolds} in the corresponding string theory backgrounds. (Such orientifolds were discussed from a bordism perspective in \cite{montero_cobordism_2021}, where they were described more generally as ``I-folds''.) We will see that supersymmetric orientifold objects are often equipped with BPS charges that generate non-torsion bordism groups, enhancing the naive $\B{Z}_2$-classes in unoriented bordism generated by $\B{RP}^i$ with the additional data of the tangential structures required by quantum gravity. 

By the BPS formula, the tensions of orientifolds are sometimes seen to carry curious minus signs. In fact, we will see that this puzzle persists even for the M-theory uplift of orientifolds, resulting in \textit{a priori} negative tension generators of bordism classes associated to orientifold backgrounds. The ECC stipulates, however, only that there \textit{exist} nonnegative-tension generators (with tension at most order 1 in Planck units). Accordingly, we expect (and we will see in examples) that each negative-tension orientifold background can always be supplemented by the addition of an appropriate number of D-branes (or M-branes) in such a way that there exists a nonnegative-tension generator carrying the topological charge associated to each bordism group.

In Secs. \ref{sssec:m-theory-orientifolds}, we study unoriented bordism in M-theory, identifying the corresponding orientifold objects which carry the associated charge and showing the existence of positive-tension generators. We then move towards Type II orientifolds in Sec. \ref{sssec:type-ii-orientifolds}, considering also cases of bordism generators with additional \textit{discrete torsion}. In the Type IIB case, we will see that these orientifolds generalize to objects carrying bordism charge for manifolds with duality bundles, which we explore briefly in Sec. \ref{sssec:nhc-s-fold}. 

\subsubsection{M-theory Orientifolds} \label{sssec:m-theory-orientifolds}

We now study the bordism classes of \textit{unoriented} manifolds in M-theory. Although we could look first at the generators of the unoriented bordism ring in Eq. \eqref{eq:unoriented-bordism}, a consistent M-theoretic tangential structure requires fermions, necessitating a $\Pin$-lift of the structure group; the relevant tangential structure is $\Pin^+$-structure \cite{mcnamara_cobordism_2019, tachikawa_why_2019}. In \cite{mcnamara_cobordism_2019}, the following bordism groups were noted: 
\begin{equation}
\begin{array}{c|ccccccccc}
k & 0 & 1 & 2 & 3 & 4 & 5 & 6 & 7 & 8 \\
\hline
\Omega^{\mathrm{Pin}^+}_k & \mathbb{Z}_2 & 0 & \mathbb{Z}_2 & \mathbb{Z}_2 & \mathbb{Z}_{16} & 0 & 0 & 0 & \mathbb{Z}_2 \times \mathbb{Z}_{32} \\
\text{Gens} & \text{pt} & - & \m{KB} & \m{KB} \times S^1_p & \mathbb{RP}^4 & - & - & - & \mathbb{HP}^2, \mathbb{RP}^8
\end{array} .
\end{equation}

As explained in \cite{mcnamara_cobordism_2019}, the topological charges of $\m{KB}$, $\m{KB} \times S^1_p$, and $\B{HP}^2$ are necessarily sourced by non-supersymmetric objects, and as such, their dynamical properties are relatively intractable. We will have nothing much to say about these objects, but note that the ECC (unsurprisingly) predicts nonnegative, sub-Planckian tensions for these objects.

We focus on the \textit{orientifold} backgrounds, carrying charges associated to $\B{RP}^k$. The manifold $\B{RP}^k$ is $\Pin^+$ precisely when $k \equiv 0, 3 \pmod{4}$. In the cases $k = 0, 4, 8$, $[\B{RP}^k]$ is \textit{unoriented} and generates torsion in $\Pin^+$-bordism (here we identify $\B{RP}^0 = \m{pt}$). For $k = 3, 7$, the classes of $[\B{RP}^3], [\B{RP}^7]$ are \textit{trivial}, so we may suspect that the corresponding orientifolds may be smoothly dissolved. This is actually not the case for $[\B{RP}^7]$, but we will see that some rather subtle M-theoretic effects are responsible for its nontriviality. 

Let us now enumerate the supersymmetric MO-planes corresponding to the bordism charges $[\B{RP}^k]$. A MO$p$-plane has transverse geometry $\B{RP}^{9-p}$; including the radial direction, we require a geometry of the form $\B{R}^{10-p}/\B{Z}_2$, where the $\B{Z}_2$ acts by involution $x \mapsto -x$. Such backgrounds preserve half supersymmetry precisely when the action of inversion $x \mapsto -x$ on fermions indeed squares to unity, which occurs precisely when $p \equiv 1, 2 \pmod{4}$. Finally, the action on the 3-form gauge field $C_3$ is fixed by the requirement of the invariance of the $C_3 \wedge G_4 \wedge G_4$ coupling, forcing $C_3 \mapsto (-1)^p C_3$. We conclude that there exist supersymmetric MO$p$-planes with transverse geometry $\B{RP}^{9-p}$ for $p = 9, 6, 5, 2, 1$. Alternatively, one can see these allowed values of $p$ by noting that they correspond exactly to the values of $k = 9-p$ for which $[\B{RP}^k]$ has $\Pin^+$-structure.

\paragraph{MO9-Plane.}

Let us first consider the class of $[\m{pt}] \in \Omega_0^{\m{Pin}^+}(\m{pt})$, which generates $\B{Z}_2$. The corresponding object carrying this topological charge is the ``MO9-plane'', which is nothing other than the usual Horava-Witten wall, with $\f{e}_8$ gauge degrees of freedom localized on its worldvolume. The length of the M-theory interval in the $E_8 \times E_8$ heterotic theory is an exact modulus, and in this sense, the Horava-Witten wall has zero tension. Wrapped on $S^1$, the Horava-Witten boundary reduces to an $\m{O8}^-$-plane in Type IIA, accompanied by 8 D8-branes to cancel the RR-charge, which also has zero tension.

In Type IIA, the $\m{O8}^- + 8 \m{D8}$ worldvolume admits deformations corresponding to sliding the D8-branes away from the $\m{O8}^-$ and thus adding sources for the Romans mass $F_0$ in the bulk. Thus, in Type IIA, we expect that a \textit{generic} O8-plane carries nonzero tension; however, the fundamental generator of the $\B{Z}_2$ class $[\m{pt}]$ in $\Omega_0^{\m{Pin}^+}(\m{pt})$ necessarily carries zero tension.\footnote{Note that unoriented backgrounds in Type IIA only make sense with an accompanying orientifold action on the \textit{worldsheet}, so $\Pin^+$-bordism does not literally make sense in Type IIA. In $\Spin$-bordism, the class $[\m{pt}]$ generates $\B{Z}$, but the associated supersymmetric trivialization (consisting of $\m{O}8^- + 8 \, \m{D} 8$) generates $\B{Z}_2$, as it must (by consistency with the M-theory lift).} The consistency of the M-theory uplift requires all of the nontrivial sources of $F_0$ to ``collapse'' onto the end-of-the-world walls, since the massive deformations of Type IIA are not expected to have an M-theoretic uplift.  

\paragraph{MO6-Plane.}

This is the direct M-theoretic lift of the O6-plane in Type IIA, which (after including the appropriate number of D6-branes) has zero tension. Like the MO9-plane, the MO6-plane has worldvolume gauge degrees of freedom, containing a 7d $\f{su}(2)$ sector \cite{hanany_orientifolds_2000}. Moreover, since the MO6 can carry no additional BPS gauge charges in M-theory, the class of the space $\B{RP}^3$ transverse to the MO6-plane must lie in the image of the map to pure $\Pin^+$-bordism. Thus, the MO6 carries trivial charge in M-theoretic bordism, meaning the MO6 configuration can be smoothly dissolved. Compare this with the M-theoretic lift of the D6-brane, which is indeed the smoothly realizable Kaluza-Klein monopole. 

\paragraph{MO5-Plane.}

We now move to the MO5-plane, which sources topological charge associated to the class of $[\B{RP}^4]$. This object was first described in \cite{dasgupta_orbifolds_1996, witten_five-branes_1996}, and was seen to carry \textit{nonzero} $G_4$-charge. We will review the discussion in the above references, computing the MO5-plane tension via its $G_4$-charge. 

The usual argument for the $G_4$-charge of the MO5-plane comes from the consistency of 6d $\C{N}=(2, 0)$ supergravity obtained from M-theory on $T^5/\B{Z}_2$. This theory arises as infinite-distance limit in the moduli space of Type IIB on K3 at which $\SL_5(\B{Z})$ discrete gauge symmetry is restored; this duality was first argued in \cite{witten_five-branes_1996}. On the other hand, the gravitational anomaly of a 6d $\C{N}=(2, 0)$ supergravity theory requires precisely 21 tensor multiplets. From the M-theory perspective, this is rather puzzling -- direct dimensional reduction suggests that there are 5 tensor multiplets in the untwisted sectors of the $T^5/\B{Z}_2$ compactification, requiring 16 additional tensor multiplets in the 32 twisted sectors corresponding to the fixed points of the $\B{Z}_2$-action on $T^5$ (with local geometry $\B{R}^5/\B{Z}_2$). 

The resolution to this puzzle, as suggested by \cite{witten_five-branes_1996}, is that one can then insert an additional 16 M5-branes transverse to the compact geometry, providing exactly the additional 16 tensor multiplets required in the 6d theory.  However, these branes each source magnetic $G_4$-flux, and thus are \textit{a priori} inconsistent on a compact background. This problem is resolved, however, if we stipulate additionally that each orbifold fixed point additionally carry magnetic $G_4$-charge: $G_4 = -\frac{1}{2}$. 

We present now a more direct argument for the M5-charge of an MO5-orientifold carrying bordism charge $[\B{RP}^4]$ using the worldvolume fermion anomaly on the M2-brane \cite{tachikawa_why_2019, montero_cobordism_2021}. A generic $\frac{1}{2}$-BPS M2-brane preserves 3d $\C{N}=8$ supersymmetry. A single fermion over $\B{RP}^4$ suffers an anomaly detected by the Atiyah-Patodi-Singer (APS) $\eta$-invariant for the $\Pin^+$ Dirac operator: $\eta_{\m{D}}(\B{RP}^4) = \pm \frac{1}{16}$, so the 3d $\C{N}=8$ M2 worldvolume theory therefore has total $\eta$-invariant $\eta_{\m{M2}}(\B{RP}^4) = \pm \frac{1}{2}$. Accordingly the M2 partition function carries an anomalous phase $\exp(i 2\pi \eta_{\m{M2}}(\B{RP}^4))$, which must be canceled by the coupling to the $C_3$-field. This coupling only cancels the phase provided that we prescribe fractional $G_4$-flux to the $\B{RP}^4$ in which the M2 propagates, yielding at least $G_4 = \mp \frac{1}{2}$. (Note that we have not been careful about the overall sign of the $\eta$-invariant, but a fully precise analysis of the anomalous worldvolume fields on the M2 will yield the appropriate sign $G_4 = -\frac{1}{2}$.)

We summarize by noting that the torsion class $[\B{RP}^4] \in \Omega_4^{\Pin^+}(\m{pt})$ in pure $\Pin^+$-bordism uplifts to an \textit{integer-valued} class $[\B{RP}^4, G_4 = -\frac{1}{2}] \in \Omega_4^{\m{M2}}(\m{pt})$ in what we have termed ``M2-bordism''. The groups $\Omega_k^{\m{M2}}(\m{pt})$ give the bordism classes of $\Pin^+$-manifolds equipped with (possibly fractional) $G_4$-flux to ensure a consistent with a probe worldvolume M2. (Importantly, note that we cannot literally interpret $\B{RP}^4$ as carrying half-integral $G_4$-flux, since $H^4(\B{RP}^4, \B{Z}) = 0$. This notation is merely meant to emphasize that the MO5-plane carries bordism charge that also detects $Q = -\frac{1}{2}$ worth of M5-brane charge. A proper geometric interpretation of this charge should be visible in full ``M-theoretic tangential structure'', whatever that may be.) By the BPS condition, we see that the MO5-planes therefore carry tension $T_{\m{MO5}}$ given by 
\begin{equation} T_{\m{MO5}} = -\frac{1}{2} T_{\m{M5}}. \end{equation}
In particular, this tension is \textit{nonzero} and \textit{negative}.

This represents an apparent conflict with the ECC (which stipulates that generators have nonnegative tension), but note simply that stacking a single M5-brane on top of an MO5 creates a defect generator carrying bordism charge associated to 
\begin{equation} [S^4, G_4 = 1] + [\B{RP}^4, G_4 = -\tfrac{1}{2}] \simeq [\B{RP}^4, G_4 = +\tfrac{1}{2}]. \end{equation}
Here, we have used the fact that $[S^4] + [M^4] \simeq [S^4 \# M^4] \simeq [M^4]$ in plain geometric bordism, and the auxiliary $G_4$-charges in the M-theoretic enhancement of the tangential structure should add under the bordism group operation. 

We note that $[\B{RP}^4, G_4 = + \frac{1}{2}]$ and $[S^4, G_4 = 1]$ equally well generate $\Omega_4^{\m{M2}}(\m{pt})$ as do $[\B{RP}^4, G_4 = - \frac{1}{2}]$ and $[S^4, G_4 = 1]$, subject to the relation that 
\begin{equation} 32 [\B{RP}^4, G_4 = - \tfrac{1}{2}] + 16 [S^4, G_4 = 1] = 32 [\B{RP}^4, G_4 = + \tfrac{1}{2}] - 16 [S^4, G_4 = 1] = 0,  \end{equation}
corresponding to the compact $T^5/\B{Z}_2$ background with 16 M5-branes to cancel the tadpole. Equivalently one can express the generators of $\Omega_4^{\m{MO5}}(\m{pt})$ as $[\B{RP}^4, G_4 = + \tfrac{1}{2}]$, generating $\B{Z}$, and $2[\B{RP}^4, G_4 = + \tfrac{1}{2}] - [S^4, G_4 = 1] = 2[\B{RP}^4, G_4 = - \tfrac{1}{2}] + [S^4, G_4 = 1]$, generating torsion. The associated ECC-consistent defect for the free generator $[\B{RP}^4, G_4 = +\frac{1}{2}]$ has tension
\begin{equation} T_{\m{MO5} + \m{M5}} = + \frac{1}{2} T_{\m{M5}}, \end{equation}
as required, while the tension of the torsion generator $2[\B{RP}^4, G_4 = - \tfrac{1}{2}] + [S^4, G_4 = 1]$ is
\begin{equation} T_{2\m{MO5} + \m{M5}} = 0. \end{equation}
Note that the worldvolume theory of a stack $n\m{M5} + \m{MO5}$ is a $D$-type 6d $\C{N} = (2, 0)$ SCFT with gauge algebra $\f{so}(2n)$. For $n \gtrsim 1$, the worldsheet count of degrees of freedom is indeed order 1, yielding $s \simeq 1$ for the generator of $[\B{RP}^4, G_4 = +\frac{1}{2}]$. 

This example is illustrative of a general paradigm -- the ECC does not predict that \textit{all} generators have nonnegative tension; it merely argues that there \textit{exist} nonnegative-tension bordism generators. An ECC-consistent set of bordism generators can be considered as representing the elementary objects in the theory, and we emphasize that the $\m{MO}5$ background is not any ``more fundamental'' than the $\m{MO5} + \m{M5}$, as both are expressible as generators for the full bordism group. In the BPS case, these nonnegative-tension generators have nonnegative BPS central charge, and are thus the appropriate objects whose worldvolume entropy can be calculated (and must be order 1, as per Conjecture \ref{conj:ecc}). 

\paragraph{MO2-Plane.}

Let us now discuss the ``electric-magnetic dual'' of the MO5-plane --- the MO2-plane. In pure $\Pin^+$-bordism, the class of the transverse manifold $\B{RP}^7$ is trivial. However, unlike the MO6-plane, we will see that consistency of with the full M-theoretic tangential structure demands that the MO2 must additionally carry (fractional) \textit{electric} M2-brane charge, corresponding therefore to a nontrivial bordism charge for the associated transverse $\B{RP}^7$ in full M-theoretic tangential structure. 

Consider a transverse $\B{RP}^7$ about a dimension-3 defect locus. We wish to calculated the associated M2-brane charge carried by this object. For this, we note the following coupling in the M-theory action: 
\begin{equation} S \supset -\int_{X_{11}} C_3 \wedge I_8(R), \end{equation}
where $I_8(R)$ is the gravitational anomaly polynomial. Note that, over a compact 8-manifold $Y_8$, we have
\begin{equation} -\int_{Y_8} I_8(R) = -\frac{\chi}{24}. \end{equation}
The cone over $\B{RP}^7$ locally looks like one of the 256 $\B{R}^8/\B{Z}_2$ singularities of $T^8/\B{Z}_2$, which has Euler characteristic $\chi(T^8/\B{Z}_2) = 384$. Accordingly, 
\begin{equation} -\int_{\B{R}^{2, 1} \times \B{R}^8/\B{Z}_2} C_3 \wedge I_8(R) = -\frac{1}{16} \int_{\B{R}^{2, 1}} C_3. \label{eq:mo2-charge}  \end{equation}
We conclude that the $G_7$-flux (which defects the M2-brane charge) of what we will call the `$\m{MO2}^-$-plane is $Q = -\frac{1}{16}$. 

In fact, the M2-charge of an $\m{MO2}^-$ is argued via M5-worldvolume anomaly cancellation, much like the magnetic dual MO5 case discussed earlier. We have seen that $\B{RP}^7$ is an \textit{a priori} consistent background for 11d supergravity (and is in fact trivial in $\Pin^+$-bordism). However, the full M-theoretic consistency requires a consistent worldvolume theory for M5-branes propagating over a background spacetime. Over $\B{RP}^7$, the M5 partition function picks up a phase $\exp(i 2\pi \,  \eta_{\m{M5}}(\B{RP}^7) )$, where $\eta_{\m{M5}}$ is the Atiyah-Patodi-Singer (APS) $\eta$-invariant. By the APS index theorem,  
\begin{equation} \eta_{\m{M5}}(X) \equiv - \int_{X} \omega_7(R) \pmod{1}, \end{equation}
where $\omega_7(R)$ is the gravitational Chern-Simons form with $\d \omega_7(R) = I_8(R)$. The appropriate anomaly cancellation condition therefore yields a Bianchi identity for $G_7$ as follows: 
\begin{equation}
    \d G_7 = I_8(R) + \cdots, \label{eq:G7-bianchi-identity}
\end{equation}
where the $\cdots$ include a $G_4 \wedge G_4$ term arising from the $C_3 \wedge C_4 \wedge C_4$ coupling in the 11d action.

Accordingly, we must consider bordism of manifolds respecting the Bianchi identity \eqref{eq:G7-bianchi-identity}, and it is precisely this condition which defines a tangential structure whose bordism theory we have called ``M5-bordism''. Compare, for instance, with the Bianchi identity for heterotic worldsheet anomaly cancellation that expresses tangential (twisted) String structure. In M5-bordism, manifolds can carry fractional $G_7$-flux given by $\eta_{\m{M5}}$, and for the class of $\B{RP}^7$ we compute $\eta_{\m{M5}}(\B{RP}^7)$, and thus the associated $G_7$-charge, as $G_7 = -\frac{1}{16}$ via \eqref{eq:G7-bianchi-identity} and \eqref{eq:mo2-charge}. Thus, the class $\B{RP}^7$, trivial in $\Pin^+$-bordism, uplifts to a class $[\B{RP}^7, G_{16} = -\frac{1}{16}]$ in M5-bordism. Unlike the MO5, however, we note that $[\B{RP}^7, G_{16} = -\frac{1}{16}]$ is \textit{torsion}, generating $\B{Z}_{256}$: the compact $T^8/\B{Z}_2$ background yields
\begin{equation} 256[\B{RP}^7, G_{16} = -\tfrac{1}{16}] = 0. \end{equation}
We emphasize that no additional M2 branes are needed to cancel the tadpole, as the $T^8/\B{Z}_2$ carries M2-charge by itself through the coupling $C_3 \wedge I_8(R)$. In fact, the charge of any codimension-$4k$ brane may be ``absorbed'' gravitationally through a Green-Schwarz-type coupling to $I_{4k}(R)$, making possible tensionful BPS codimension-$4k$ torsion defects like we have just observed. Compare also with the case of an NS5-brane in heterotic string theory, which is BPS and has positive tension, generating $\B{Z}_{24} = \Omega^{\String}_3(\m{pt})$.

By the BPS formula, the MO$2^-$ tension $T_{\m{MO2}^-}$ is 
\begin{equation} T_{\m{MO2}^-} = -\frac{1}{16} T_{\m{M2}}, \end{equation}
which is negative. However, in M5-bordism, we have generators $[\B{RP}^7, G_7 = -\frac{1}{16}]$ and $[S^7, G_7 = 1]$. From these, we easily deduce the appropriate ECC-consistent combinations. Simply take 15 $\m{MO2}^-$-planes and stack an M2, which carries charge 
\begin{equation} 15[\B{RP}^7, G_7 = -\tfrac{1}{16}] + [S^7, G_7 = 1]. \end{equation}
The associated ECC-consistent BPS defect has tension
\begin{equation} T_{15 \m{MO2}^- + \m{M2}} = + \frac{1}{16} T_{\m{M2}}. \end{equation}
The classes $15[\B{RP}^7, G_7 = -\tfrac{1}{16}] + [S^7, G_7 = 1]$ and $[S^7, G_7 = 1]$ generate $\Omega_7^{\m{M5}}(\m{pt})$, and each have nonnegative tension.\footnote{It may seem unnatural that $15[\B{RP}^7, G_7 = -\tfrac{1}{16}] + [S^7, G_7 = 1]$ and $[S^7, G_7 = 1]$ are both apparently non-torsion generators of a group with a torsion factor $\B{Z}_{256}$ of $\Omega^{\m{M5}}_7(\m{pt})$ generated by $[\B{RP}^7, G_7 = -\frac{1}{16}]$. In fact, both are secretly torsion! The class $[S^7, G_7 = 1] \in \Omega^{\m{M5}}_7(\m{pt})$ generates $\B{Z}_{30}$, and a nullbordism of $30 [S^7, G_7 = 1]$ is given by the so-called \textit{Bott manifold} \cite{freed_consistency_2021}, which will be discussed in further detail in \cite{AlvarezGarcia-heterotic-202x}.}

We note briefly that the $\m{MO2}^-$ is not the only possible MO2-plane, however, as there is a possible M-theoretic discrete torsion arising from $G_4$-flux wrapping a 4-cycle in $\B{RP}^7$, valued in $H^4(\B{RP}^7, \B{Z}) = \B{Z}_2$. The corresponding $C_3$-charge is computed by dimensional reduction of the M-theory $C_3 \wedge G_4 \wedge G_4$ term, and is found to be 
\begin{equation} Q =  \int_{\B{RP}^7} \frac{1}{2}  C_3 \wedge G_4 = \frac{1}{4}.  \end{equation}
In the presence of this discrete torsion, we may call the corresponding object a $\m{MO}2^+$-plane, and its $C_3$-charge is $Q = + \frac{3}{16}$, and so the MO$2^+$ tension $T_{\m{MO2}^+}$ is
\begin{equation} T_{\m{MO2}^+} = \frac{3}{16} T_{\m{M2}}, \end{equation}
which is already positive and ECC-consistent.

\paragraph{MO1-Plane.}

We now study the MO1-plane, sourcing topological charge associated to the generator $[\B{RP}^8] \in \Omega^{\Pin^+}_8(\m{pt}) = \B{Z}_{32}$. There are no BPS charges that this object can carry in M-theory, and correspondingly, we expect that it remains a BPS torsion generator in full M-theoretic structure. The MO1-plane is the M-theoretic uplift of the O0-planes in Type IIA, and its mass can be derived from the mass of the corresponding O0s. From a purely M-theoretic perspective, the MO1-plane arises from a decompactification of the O0-plane wrapped along the M-theory circle, and corresponds to M-theory on $\B{R}^9/\B{Z}_2$. It carries on its worldvolume a single chiral fermion \cite{hanany_orientifolds_2000}. 

As with higher O$p$-planes the O$0$-plane carries various discrete torsion classes on its transverse $\B{RP}^8$ in the Type IIA description. The variant with zero discrete torsion is typically called the O$0^-$-plane, and as is well-known \cite{tachikawa_why_2019}, the $\m{O}0^-$-plane carries fractional D0-brane charge $Q_{\m{O0}} = -\frac{1}{16}$, so that $T_{\m{O}0^-} = -\frac{1}{16} T_{\m{D}0}$. Wrapping the MO1 on a circle requires a choice of boundary conditions for the worldvolume fermion. For the NS (anti-periodic) boundary conditions, we obtain an $\m{O}0^-$ from a wrapped $\m{MO}1$ \cite{hanany_orientifolds_2000}, and $T_{\m{MO}1} R_{11} = T_{\m{O}0}$. Accordingly, in 11-dimensional Planck units, we find that $T_{\m{MO}1} \simeq 1/R_{11}^2$, which indeed vanishes in the M-theoretic limit. Of course, the supersymmetric MO1 must be tensionless, as it can carry no M-theoretic BPS charge.  

\subsubsection{Type II Orientifolds and Discrete Torsion} \label{sssec:type-ii-orientifolds}

In this section, we move away from purely geometric MO-orientifolds to study perturbative orientifold backgrounds in Type II string theory. An O$p$-plane locally looks like a copy of $\B{R}^{k}/\B{Z}_2$ for $k = 9-p$, where the $\B{Z}_2$ acts by $x \mapsto -x$ accompanied by an action by $\Omega \cdot (-1)^{\m{F_L}}$ on the string worldsheet.

The precise bordism charge carried by orientifold backgrounds was described in \cite{tachikawa_why_2019, montero_cobordism_2021}. Briefly, in Type IIA, BPS orientifolds carry \textit{unoriented} bordism charge valued in either $\Pin^+$ or $\Pin^-$ bordism, depending on whether the orientifold background carries an even or odd number of reflections along the M-theory circle. In Type IIB, orientifolds source \textit{oriented} bordism charge equipped with a (possibly Spin-lifted) $\B{Z}_2$ duality bundle within the central $\B{Z}_4$ of $\m{Mp}(2, \B{Z})$. The bordism group in which the $[\B{RP}^{8-p}]$ bordism charge carried by an orientifold O$p$-plane is valued is summarized in Table \ref{tab:orientifold-bordism-charges}.

\begin{table}[h]
    \centering
    \begin{tabular}{c|c|c|c|c}
        Dimension $k$ & $4n$ & $4n + 1$ & $4n + 2$ & $4n + 3$  \\
        \hline
        Bordism Group & $\Omega_k^{\Pin^+}(\m{pt})$ & $\Omega_k^{\Spin\text{-}\B{Z}_4}(\m{pt})$ & $\Omega_k^{\Pin^-}(\m{pt})$ & $\Omega_k^{\Spin}(B\B{Z}_2)$ 
    \end{tabular}
    \caption{The bordism groups in which orientifold bordism charges $[\B{RP}^k]$ in various dimenions are valued.}
    \label{tab:orientifold-bordism-charges}
\end{table}

There is an additional consistency condition on the RR charge $F_{8-p}$ for orientifold O$p$-plane backgrounds that arises from the anomaly cancellation condition for D$(6-p)$-branes on the transverse $\B{RP}^{8-p}$, analogous to the M2/M5 worldvolume anomaly cancellation condition that defined tangential M2/M5-structure for MO5/MO2-plane backgrounds in Sec. \ref{sssec:m-theory-orientifolds}. This condition was described in \cite{tachikawa_why_2019} in the case of O$p^+$-plane backgrounds with discrete torsion (which we will describe later) and pertains to the $\eta$-invariant of the D$(6-p)$-brane worldvolume fermions. For O$p^-$-plane background without discrete torsion, there is an additional contribution  to the $\eta$-invariant due to the D$(6-p)$-brane worldvolume gauge fields \cite{tachikawa_why_2019, hsieh_anomaly_2019}, with the net result that $\eta_{\m{D}(6-p)}(\B{RP}^{8-p}) = -2^{5-p}$. Accordingly, we expect an O$p^-$-plane to carry negative RR-charge $F_{8-p} = -2^{5-p}$, which is fractional for $p < 5$, corresponding finally to a class $[\B{RP}^{8-p}, F_{8-p} = -2^{5-p}] \in \Omega^{\m{D}(6-p)}_{8-p}(\m{pt})$ in ``D$(6-p)$-brane bordism''. 

\paragraph{$\m{O}p$-Planes for $p \geq 5$.} 

Let us now study the generators of bordism charge curresponding to orientifold $\m{O}p$-plane backgrounds for $p \geq 5$, where there is no charge fractionalization. For $p \geq 5$, $Q_{\m{O}p^-}$ is an integer multiple of $Q_{\m{Dp}}$. Accordingly, adding an integer number of D$p$-branes on top of an O$p^-$-plane produces a background carrying no net RR-flux. We see that the combination $\m{O}p^- + 2^{p-5} \m{D}p$ carries bordism charge 
\begin{equation} [\B{RP}^{8-p}, F_{8-p} = -2^{p-5}] + 2^{5-p}[S^{8-p}, F_{8-p} = 1] , \end{equation} generating torsion. This, along with the ordinary D-brane class $[S^{8-p}, F_{8-p} = 1]$, generate $\Omega_{8-p}^{\m{D}}(\m{pt})$.

The tension of the torsion $\m{O}p^- + 2^{p-5} \m{D}p$ configuration is clear. For $5 \leq p < 7$, the tension is exactly zero by the BPS formula. For $p = 7$, the BPS tension also appears to be zero. However, the configuration of $\m{O}7^- + 4 \m{D}7$ has a nonzero deficit angle $\delta = \pi$, admitting a decomposition into six (not mutually local) D7-branes. Correspondingly, it generates the $\Spin$-lift $\B{Z}_4 = \Omega^{\Spin\text{-}\B{Z}_4}_1(\m{pt})$ and four such $\m{O}7^- + 4 \m{D}7$ configurations can be consistently trivialized over the $T^2/\B{Z}_2$ ``pillowcase'' geometry, which looks like $\B{P}^1$ with four conical deficit singularities with angle $\delta = \pi$. The $\m{O}7^- + 4 \m{D}7$ induces no varying axiodilaton profile, so its total tension is precisely $T \simeq 2\delta = 2\pi$ in Planck units, in agreement with the ECC. Finally, for the $\m{O} 8^- + 8 \m{D}8$ configuration, as a domain wall, can also be understood in a sense to have zero tension since the Type I' interval length is an exact modulus.

\paragraph{$\m{O}p$-Planes for $p < 5$.}

For $p < 5$, the situation is rather different due to the \textit{fractionalization} of $\m{O}p^-$-plane charges. We find that the O$p^-$ planes are \textit{a priori} negative tension generators for the appropriate bordism groups; however, reasoning essentially analogous to the discussion in Sec. \ref{sssec:m-theory-orientifolds} for the MO5- and MO2-planes will argue that there exist positive-tension bordism generators for the same groups. Indeed, we note that the classes of $[\B{RP}^{8-p}, -2^{5-p}]$ and $[S^{8-p}, 1]$, corresponding respectively to an O$p^-$-plane and a D$p$-brane, generate $\Omega^{\m{D}(6-p)}_{8-p}(\m{pt})$. The compact $T^{9-p}/\B{Z}_2$ background for Type II yields a condition 
\begin{equation} 
2^{9-p} [\B{RP}^{8-p}, F_{8-p} = -2^{5-p}] +  2^{13-2p} [S^{8-p}, F_{8-p} = 1] = 0. 
\end{equation}
Analogously to the MO2- and MO5-examples, we choose generators $(2^{p-5} - 1) [\B{RP}^{8-p}, -2^{5-p}] + [S^{8-p}, 1]$, generating $\B{Z}$, and $2^{p-5} [\B{RP}^{8-p}, -2^{5-p}] + [S^{8-p}, 1]$, generating torsion. The tension of the integer-valued generator is 
\begin{equation} T_{(2^{5-p} - 1) \m{O}p^- + \m{D}p} = + 2^{5-p}, \end{equation}
and the tension of the torsion generator is zero. 

\paragraph{Discrete Torsion.}
We now discuss variants of the O$p^-$-planes considered earlier that carry additional topological charges. The transverse spaces $[\B{RP}^{8-p}]$ may be equipped with (higher) gauge bundles associated to the RR and NSNS fluxes and connections in the corresponding theory, and these bundles further modify the D-brane charge of the orientifold. Accordingly, we can study different classes of O$p$-planes carrying different bordism charges that are labeled by their discrete torsion.

Let us focus on the example of the O3-plane. The orientifold acts on the NSNS 2-form connection via $B_2 \mapsto -B_2$, and by $\SL_2(\B{Z})$-invariance (recall, for instance, that the O7-plane is implemented by the central $\B{Z}_2$ generator of $\SL_2(\B{Z})$), the action on the RR 2-form connection $C_2$ is via $C_2 \mapsto -C_2$. Accordingly, the classes $B_2, C_2$ are valued in \textit{twisted} (differential) cohomology with integer coefficients over $\B{RP}^5$, and are classified up to topological equivalence via the corresponding fluxes $H_3, F_3 \in H^3(\B{RP}^5, \tilde{\B{Z}})$, where $\tilde{\B{Z}}$ represents a $\B{Z}$-bundle over $\B{RP}^5$ twisted by a $\B{Z}_2$-action on the fiber of the pullback $\B{Z}$-bundle over $S^5$ along the twofold cover $S^5 \onto \B{RP}^5$. 

It is easily computed that $H^3(\B{RP}^5, \tilde{\B{Z}}) = \B{Z}_2$, and correspondingly, there are four different types of O3-planes for each associated discrete torsion class. Generalizing to O$p$-planes for $p < 6$, it is seen that in addition to the NSNS discrete torsion associated to a class $[H_3] \in H^3(\B{RP}^{8-p}, \tilde{\B{Z}}) = \B{Z}_2$, there is a RR discrete torsion class $[F_{6-p}] \in H^{6-p}(\B{RP}^{8-p}, \tilde{\B{Z}}) = \B{Z}_2$. Accordingly, there are four different O$p$-plane variants for all $p < 6$. Corresponding to each O$p$-plane is an associated effective RR-charge, which is in the O$p^+$ case computed from the $\eta$-invariant for the corresponding D$(6-p)$-brane worldvolume in the associated discrete torsion background \cite{tachikawa_why_2019, hsieh_anomaly_2019}. This data is tabulated in Table \ref{tab:Op-planes}. 

\begin{table}[ht]
    \centering
    \begin{tabular}{c|c|c}
        Name & Discrete Torsion $(F_{6-p}, H_3)$ & RR $F_{8-p}$-charge \\
        \hline
        O$p^-$ & $(0, 0)$ & $-2^{p-5}$ \\
        O$p^+$ & $\left(0, \frac{1}{2} \right)$ & $+2^{p-5}$  \\
        $\widetilde{\m{O}p^-}$ & $\left( \frac{1}{2}, 0 \right)$  & $-2^{p-5}+ \frac{1}{2} $  \\ 
        $\widetilde{\m{O}p^+}$ & $\left( \frac{1}{2}, \frac{1}{2} \right)$ & $+2^{p-5}$  \\ 
    \end{tabular}
    \caption{D-brane charges of orienifolds with discrete torsion.}
    \label{tab:Op-planes}
\end{table}

Most of these discrete torsion variants are already positive tension generators. For the negative-tension objects, one employs a procedure similar to the analysis with the O$p^-$- and MO-plane backgrounds to obtain positive-tension generators. 

\subsubsection{Non-Higgsable Clusters and S-folds} \label{sssec:nhc-s-fold}

In this section, we begin with a consideration of other defects carrying torsion bordism charge, generalizing the orientifold objects considered previously. A natural candidate for supersymmetric generalizations of orientifolds are (singular) \textit{non-Higgsable clusters} (NHCs) and \textit{S-folds} \cite{aharony_s-folds_2016}, both of which correspond to singular backgrounds of the form $\B{C}^n/\B{Z}_k$ coupled with an action by a $\B{Z}_k$ subgroup of the non-perturbative duality group, realized as an enhanced gauge symmetry at an interior point of the moduli space. As such, NHCs and S-folds are ``frozen'' at interior points in the moduli space and cannot have moduli-dependent tension, so the ECC predicts that their tension be sub-Planckian. 

In the case of NHCs, we will see that these supersymmetric objects cannot carry BPS charges, and accordingly, their tensions must be zero. S-folds, however, can carry D3-brane charge, and we will see that, much like the case of orientifolds with discrete torsion, they carry \textit{fractional} D3-brane charges. This provides yet another example in which bordism classes are \textit{stipulated} to carry fractional charge by the overall consistency of the theory, and the associated BPS defects have Planckian tension. 

At a first pass, NHCs and S-folds are detected by bordism groups $\Omega_k^{\Spin\text{-}\m{Mp}(2, \B{Z})}(\m{pt})$ associated to Type IIB with a ($\Spin$-lifted) $\SL(2, \B{Z})$ duality bundle. We already studied the sevenbranes carrying charge in $\Omega_1^{\Spin\text{-}\m{Mp}(2, \B{Z})}(\m{pt}) = \B{Z}_{24}$ in Sec. \ref{ssec:codimension-leq-2}; in this section, we focus on the groups $\Omega_3^{\Spin\text{-}\m{Mp}(2, \B{Z})}(\m{pt})$ and $\Omega_5^{\Spin\text{-}\m{Mp}(2, \B{Z})}(\m{pt})$, computed first in \cite{debray_chronicles_2023}:
\begin{equation} \Omega_3^{\Spin\text{-}\m{Mp}(2, \B{Z})}(\m{pt}) = \B{Z}_2 \oplus \B{Z}_3 \end{equation}
\begin{equation} \Omega_5^{\Spin\text{-}\m{Mp}(2, \B{Z})}(\m{pt}) = \B{Z}_2 \oplus \B{Z}_{32} \oplus \B{Z}_9 \end{equation}
The bordism charges associated to these groups are represented by \textit{lens spaces} of the form $L^{2k+1}_n = S^{2k+1}/\B{Z}_n$, defined as the link of the origin about a $\B{C}^k/\B{Z}_n$ orbifold singularity. The defects generating these bordism groups are in some sense non-perturbative generalizations of Type IIB orientifolds, and may be treated similarly.

\paragraph{Non-Higgsable Clusters.} The generators of the charge $\Omega_3^{\Spin\text{-}\m{Mp}(2, \B{Z})}(\m{pt}) = \B{Z}_2 \oplus \B{Z}_3$ are visible in F-theory as (singular limits of) \textit{Non-Higgsable Clusters} (NHCs). The associated generators of $\B{Z}_2$ and $\B{Z}_3$ are the lens spaces $[L^3_4]$ and $[L^3_3]$, respectively, associated to certain orbifold singularities $\B{C}^2/\B{Z}_4$ and $\B{C}^2/\B{Z}_3$ with corresponding $\B{Z}_4$ and $\B{Z}_3$ duality bundles, where the $\B{Z}_4$ and $\B{Z}_3$ actions are generated by the $\Spin$-lifts of $\SL(2, \B{Z})$ generators.

Let us focus on the singular NHC generating $[L^3_3]$. As described in \cite{debray_chronicles_2023}, this corresponds to F-theory on the non-compact background $(\B{C}^2 \times T^2)/\B{Z}_3$, where $\B{Z}_3$ acts on $C^2 \times T^2$ as $(z_1, z_2, U^4 \cdot \tau) \mapsto (\omega z_1, \omega z_2, \omega \tau)$ and $\tau$ is the complex-structure parameter of the F-theory $T^2$ and $U^4$ generates the $\B{Z}_3$ duality action. This NHC corresponds to a curve in the base $B$ of an elliptically-fibered Calabi-Yau threefold with self-intersection $-3$ and supports a $\f{su}(3)$ gauge algebra which admits no Higgs-branch deformations. The associated 6d object preserves $\C{N} = (1, 0)$ supersymmetry but admits no BPS charges in the Type IIB background, since the action of $\B{Z}_3$ on the fluxes $(F_3. H_3)$ is nontrivial. Accordingly, the singular NHC generating $[L^3_3]$ is exactly tensionless, as a BPS torsion bordism generator. The exact same reasoning argues that the singular NHC generating $[L^3_4]$ is tensionless. Note that $2 [L^3_4]$, twice the $\B{Z}_4$ generator in $\Omega_5^{\Spin\text{-}\m{Mp}(2, \B{Z})}(\m{pt})$, is sourced by the (tensionless) $\m{O}5^- + \m{D}5$ described in Sec. \ref{sssec:type-ii-orientifolds}. 

\paragraph{S-Folds.} We next consider the group $\Omega_5^{\Spin\text{-}\m{Mp}(2, \B{Z})}(\m{pt}) = \B{Z}_2 \oplus \B{Z}_{32} \oplus \B{Z}_9$. The associated defects that carry these bordism charges are called \textit{S-folds}, and a stack of D3-branes probing an S-fold has on its worldvolume a strongly coupled 4d $\C{N}=3$ superconformal theory \cite{aharony_s-folds_2016, garcia-etxebarria_n3_2016}. Unlike NHCs and like their orientifold cousins from Sec. \ref{sssec:type-ii-orientifolds}, S-folds carry D3-brane charge, which may be negative.

The S-fold backgrounds that generate $\Omega_5^{\Spin\text{-}\m{Mp}(2, \B{Z})}(\m{pt})$ were described in $\cite{debray_chronicles_2023}$, and the associated bordism charges are given by lens spaces $[L^5_3]$, $[L^5_4]$, and $[\tilde{L}^5_4]$, (where $[\tilde{L}^5_4]$ differs from $[L^5_4]$ by the certain equivariant $\eta$-invariants, see Appendix C of \cite{debray_chronicles_2023} for details) with $\B{Z}_3$ or $\B{Z}_4$ duality bundles. The D3-brane charge of S-folds may be argued from an entropic perspective \cite{aharony_s-folds_2016}. For simplicity, let us concentrate on S-fold background corresponding to $(\B{C}^3 \times T^2)/\B{Z}_3$, where $\B{Z}_3$ acts via $(z_1, z_2, z_3, \tau) \mapsto (\omega z_1, \omega z_2, \omega z_3, U^4 \cdot \tau)$, which generates bordism charge associated to the lens space $[L^5_3]$; the other cases are analogous.

The worldvolume central charge of the 4d $\C{N}=3$ SCFT obtained by placing $N$ D3-branes on an S-fold is expected to scale as $S \sim \frac{3}{4} (N + \delta)^2$, where $\delta$ is the fractional D3-charge of the S-fold background (here the factor of $\frac{3}{4}$ arises from the $\B{Z}_3$ quotient). On the other hand, the central charges $a, c$ are related to the Coulomb branch generator operator dimensions $\Delta_i$ via 
\begin{equation} 2a - c = \frac{1}{4} \sum_i (2\Delta_i - 1). \end{equation}
Thus, finding the Coulomb branch generators of a stack of $N$ D3-branes probing the $(\B{C}^3 \times T^2)/\B{Z}_3$ singularity yields the D3-charge of this S-fold. This calculation was performed in \cite{aharony_s-folds_2016}, where it was found that $\Delta_i \in \{3, 6, \cdots, 3(N-1), N\}$, yielding
\begin{equation} S \sim 2a - c \sim \frac{1}{4} (3N^2 - 2N) \sim \frac{3}{4}\left( N - \frac{1}{3} \right)^2, \q N \gg 1   \end{equation}
so that the D3-brane charge is $F_5 = \delta = -\frac{1}{3}$. 

We see therefore there that S-fold backgrounds are generally forced to carry nontrivial D3-brane charge, which can be negative. One can in principle also understand this D3-brane charge as arising from the $\eta$-invariant of appropriate D3-worldvolume operators which must in turn be absorbed by $F_5$-flux. For instance, a single worldvolume Dirac fermion transforming with charge $q = \frac{1}{2}$ under $\Spin\text{-}\B{Z}_3$, it was computed by the equivariant APS index theorem in \cite{debray_chronicles_2023} that $\eta_{1/2}^{\m{D}} = -\frac{1}{9}$. The appropriate combination of anomalous probe worldvolume degrees of freedom should yield the D3-charge of the $\B{Z}_3$ S-fold, but to the authors' best knowledge, no such systematic analysis has been performed (such as for perturbative orientifolds in \cite{tachikawa_why_2019}). 

Nevertheless, we expect there to exist a full theory of ``D3-$\Spin\text{-}\m{Mp}(2, \B{Z})$-bordism'' with nonperturbative (Spin-lifted) duality bundle and the appropriate $F_5$-charge to ensure the consistency of probe D3 worldvolumes, and we expect the S-fold class $[L^5_3]$ to uplift to $[L^5_3, F_5 = -\frac{1}{3}] \in \Omega^{\text{D3-}\Spin\text{-}\m{Mp}(2, \B{Z})}_5(\m{pt})$. This class is a non-torsion generator in $\Omega^{\text{D3-}\Spin\text{-}\m{Mp}(2, \B{Z})}_5(\m{pt})$. Just like with the orientifold examples, however, we may equally well consider $2[L^5_3, F_5 = -\frac{1}{3}] + [S^5, F_5 = 1]$ as a BPS positive-tension non-torsion generator of $\Omega^{\text{D3-}\Spin\text{-}\m{Mp}(2, \B{Z})}_5(\m{pt})$ alongside a torsion generator $3[L^5_3, F_5 = -\frac{1}{3}] + [S^5, F_5 = 1]$. Both of these are ECC-consistent BPS generators with nonnegative tension. 

Like orientifolds, S-folds have an M-theoretic uplift corresponding to M-theory on $\B{C}^4/\B{Z}_k$ orbifold singularities. These singular backgrounds carry M2-brane charge, detected by $\eta_{\m{M5}}(L^7_k)$, the M5-brane worldvolume $\eta$-invariant of the associated lens space (analogously to the case of MO2-planes in Sec. \ref{sssec:m-theory-orientifolds}). One readily verifies that the M-theoretic S-fold backgrounds also have ECC-consistent nonnegative tension generators in M5-bordism (as we defined in Sec. \ref{sssec:m-theory-orientifolds}.)

\paragraph{S-Strings and S-Instantons.} We conclude by briefly noting that there exist analogs of S-folds in higher codimension, detected by $\Omega_7^{\Spin\text{-}\m{Mp}(2, \B{Z})}(\m{pt})$ and $\Omega_9^{\Spin\text{-}\m{Mp}(2, \B{Z})}(\m{pt})$. These objects are nonperturbative generalizations of Type IIB O1-planes and ``O$(-1)$-instantons'', which were labeled ``S-strings'' and ``S-instantons'' in \cite{debray_chronicles_2023}. We expect there to exist nonnegative-tension bordism generators analogously to the S-fold and NHC cases. Note that the \textit{nonperturbative} S-strings cannot carry BPS charges due to the nontrivial $\SL(2, \B{Z})$ action on $(F_7, H_7)$, and indeed we expect S-strings to be tensionless.

\section{Non-Supersymmetric Examples} 
\label{ssec:non-susy-heterotic-type-i}

We now move to study our first genuinely non-supersymmetric examples. In Sec. \ref{sssec:type-i-dbranes}, we consider torsion-valued $\KO$-theoretic D-brane charges in Type I, an example where heavy non-supersymmetric defects in the Type I frame become light on the heterotic side, necessitating the duality. On the heterotic side, we study in Sec. \ref{ssec:non-susy-heterotic} the nonsupersymmetric branes in \cite{kaidi_non-supersymmetric_2024, kaidi_non-supersymmetric_2023} as an important example of why \textit{bordism} is a necessary characterization of topological charges. Some of the branes predicted in \cite{kaidi_non-supersymmetric_2024} apparently have infinite tension, but the apparent contradiction with the ECC is resolved by recognizing that these objects do not independently trivialize a bordism class and are to be understood as relative defects attached to infinitely long ``flux tubes''. Finally, in Sec. \ref{ssec:implications-geometry}, we study possibly non-supersymmetric configurations obtained by D-branes wrapping cycles in Calabi-Yau compactifications, and we predict geometric constraints on the volumes of torsion middle-dimensional cycles in CY threefolds.

\subsection{Type I D-branes} \label{sssec:type-i-dbranes}

We first study $\m{KO}$-theoretic D-brane charges in Type I string theory. We will see that these defects, including the non-BPS objects associated to the torsion $\m{KO}$-theoretic classes, become light either in the perturbative Type I or heterotic regime. Looking only at the Type I perspective, where these non-BPS $\m{KO}$-theoretic charges were first understood, this is actually rather mysterious, but the duality with $\Spin(32)/\B{Z}_2$ heterotic cleanly illustrates a regime in which these objects become light. 

It is important to identify which bordism groups are trivialized by the objects of interest. In his paper \cite{witten_d-branes_1998}, Witten argues that D-brane charge be properly understood as taking values in K-theory instead of (ordinary) cohomology. In Type II theory, it is argued that D-branes should be understood in terms of their Chan-Paton bundles, given by rank-$n$ complex vector bundles, with the addition of D-branes given by direct sum of Chan-Paton bundles. The picture in Type I is analogous, the only difference being that the Chan-Paton bundles are \textit{real} instead of complex due to the presence of the background $\m{O}9^-$ plane. The topoological equivalence classes of such real vector bundles are characterized by the zeroth space of a \textit{spectrum}\footnote{A spectrum is a collection of topological spaces $E_n$ with isomorphisms $\Omega E_n \mathop{\tto}^\sim E_{n-1}$, where $\Omega E_n$ is the (based) loop space of $E_n$. The \textit{(stable) homotopy groups} of a spectrum $E$ are given by 
\begin{equation} \pi_n(E) = \lim_{k \to \infty} \pi_{n+k}(E_k). \end{equation}} called $\m{KO}$-theory, and this zeroth space $\KO_0$ is given by 
\begin{equation} \m{KO}_0 = B\m{O} \times \B{Z}. \end{equation}
The spectrum $\m{KO}$ is 8-periodic, and its homotopy groups are given as follows: 
\begin{center}
\begin{tabular}{c|c|c|c|c|c|c|c|c|c|c|c}
    $k$ & 0 & 1 & 2 & 3 & 4 & 5 & 6 & 7 & 8 & 9 & 10 \\
    \hline 
    $\pi_k(\KO)$ & $\B{Z}$ & $\B{Z}_2$ & $\B{Z}_2$ & $\cdot$ & $\B{Z}$ & $\cdot$ & $\cdot$ & $\cdot$ & $\B{Z}$ & $\B{Z}_2$ & $\B{Z}_2$ 
\end{tabular}
\end{center}
with $\pi_{k+8}(\KO) = \pi_k(\KO)$. Since Type I D-branes are ``stable'' versions of $O(n)$ gauge bundle configurations, we find that each homotopy group $\pi_k(\KO)$ is therefore a Type I D$(9-k)$-brane configuration which realizes the associated (stable) gauge instanton configuration supported on the compactified $S^k$ of the transverse $\B{R}^k$. 

Accordingly, the ``charge'' sourced by each defect in dimension $(10-k)$ that is detected magnetically by the linking $S^{k-1}$ is associated to the bordism class of $[S^{k-1}]$ equipped with a class degree-$k$ $\KO$-theory. This class is therefore mathematically represented by the following bordism group generators: 
\begin{equation} [S^{k-1}_{\Sigma^{-1} \KO}] \in \Omega^{\chi}_{k-1}(\Sigma^{-1} \KO),  \end{equation}
where $\Sigma^{-1}\KO$ is the degree-1 shift of $\KO$-theory with the property that $\pi_{k-1}(\Sigma^{-1}\KO) = \pi_k(\KO)$ and $\chi$ is an appropriate (stable) tangential structure for Type I/Heterotic (we will revisit the characterization of the precise tangential structure in the next section). 

Let us now examine the bordism charges associated to these classes and how they appear in Type I and the dual heterotic theory. The integer-valued classes associated to $k = 0, 4, 8$ are sourced by nothing other than the (BPS) Type I D9, D5, and D1-branes, respectively. The tension of the background O9-plane (along with the accompanying 32 D9-branes) is zero due to the vanishing cosmological constant. The Type I D5-brane, being nothing but an orientifold projections of the corresponding Type II object, becomes asymptotically light in the Type I regime, being the counterpart of the (heavy) heterotic NS5-brane. The Type I D-string, on the other hand, becomes light in the heterotic limit, being dual to nothing other than the fundamental heterotic string. 

The torsion classes, on the other hand, are more interesting, as they represent genuine non-BPS D-brane states of the Type I theory, and they provide a first foray into the use of our conjecture to understand the non-BPS spectrum of the theory. 

The $\B{Z}_2$-classes associated to $k = 1, 2$ are expected to be non-BPS torsion eight- and sevenbranes, respectively. By duality with the heterotic side, the global form of the Type I gauge group becomes $\Spin(32)/\B{Z}_2$, so the eightbrane is killed by the uplift from the $\m{O}(32)$ to $\Spin(32)/\B{Z}_2$, since $\pi_1(B\Spin(32)/\B{Z}_2) = 0$. The charge associated to $\pi_2(\KO)$, on the other hand, is slightly different. About the Type I vacuum, $\mO(32)$ is lifted to $\Spin(32)/\B{Z}_2$, and the group $\pi_2(B\Spin(32)/\B{Z}_2) = \B{Z}_2$ remains nonzero. However, it no longer describes a $\KO$-theoretic charge like $\pi_2(\KO)$ to which we would associate a genuine Type I D-brane. Indeed, we have seen that D-branes are nothing but topological classes of (virtual) real Chan-Paton bundles, which are canonically identified with their principal frame $O(n)$-bundles. Passing to the $\Spin$-lift of an $O(n)$-bundle leaves behind its identification with the associated real rank-$n$ Chan-Paton bundle, and thus its identification with D-brane charge. We conclude that $\pi_2(B\Spin(32)/\B{Z}_2) = \B{Z}_2$ cannot label a ``Type I D7-brane'' in the usual sense.

Nevertheless, this $\B{Z}_2$ is a genuine topological charge, and one may still consider the bordism class of $[S^2]$ with such a nontrivial $\Spin(32)/\B{Z}_2$ bundle and ask about its (singular) trivialization. This object will be a \textit{six}brane, and is described from the dual heterotic perspective in \cite{kaidi_non-supersymmetric_2024, kaidi_non-supersymmetric_2023} (and later in Sec. \ref{ssec:non-susy-heterotic}). We do not presently know how to characterize the tension of this sixbrane, but like the D5-brane, we expect that it becomes light in the Type I regime itself. For further discussion, see the end of Sec. \ref{ssec:non-susy-heterotic}.

We focus now on the states for $k = 8, 9$ carrying genuine $\KO$-theoretic charge about the Type I vacuum, associated to the bordism generators $[S^8_{\Sigma^{-1} \KO}]$ and $[S^9_{\Sigma^{-1} \KO}]$. These objects have indeed been well-understood since \cite{witten_d-branes_1998}. However, we emphasize the way in which these objects support our conjecture, which is especially important due to the dearth of tractable non-BPS examples.

\paragraph{Type I Gauge Spinor.} The bordism charge $[S^8_{\Sigma^{-1}\KO}]$ is carried by a particle configuration with a (stable) gauge bundle with nontrivial $\pi_9(\KO)$. It was argued in \cite{witten_d-branes_1998} that this particle, carrying $\B{Z}_2$-valued topological charge, transform in a spinor representation of the gauge bundle. Briefly, for such a particle, corresponding to a $\KO$-theoretic charge over $S^9$, we may associate a rank-$n$ real vector bundle (or $SO(n)$ gauge bundle) $U$ along with a trivial $SO(m)$ bundle $W$ corresponding to the remaining D-branes which do not contain the particle in their worldvolume gauge theory. In the Type I background, the fermionic gluinos transform in the adjoint of $U \oplus W$, which contains $U \otimes W$ as an irreducible subrepresentation under $SO(n) \times SO(m)$. The zero modes of the Dirac operator twisted by $U$, of which there are an odd number, correspond to $SO(m)$-representations valued in fermionic degrees of freedom, which must therefore uplift to representations of $\Spin(m)$. We conclude that the trivialization of the class $[S^8_{\Sigma^{-1}\KO}]$ must be a massive object transforming in a spinor representation of the gauge algebra associated to the $\KO$-theoretic gauge bundle. 

For the Type I vacuum with gauge algebra $\f{so}(32)$, we see that the non-perturbative Type I gauge group must uplift to a Spin cover of $\SO(32)$, which, as we have seen, is nothing other than $\Spin(32)/\B{Z}_2$.\footnote{Note that the uplift to $\Spin(32)/\B{Z}_2$, and not $\Spin(32)$, 
may be understood via $\pi_8(\Sigma^{-1} \KO) = \B{Z}_2$ --- were the uplift to be instead $\Spin(32)$, we would expect a $(\B{Z}_2 \times \B{Z}_2)$-valued topological charge instead, corresponding to the structure of the weight lattice of $\Spin(32)$. SR thanks Ivano Basile for discussions about this point.} On the heterotic side, the associated massive spinors are nothing other than the massive heterotic string modes populating the weight lattice $D_{16}^+$ of $\Spin(32)/\B{Z}_2$, which become light relative to Planck in the perturbative heterotic string limit. Our conjecture, viewed in this light, essentially ``post-dicts'' the existence of the heterotic limit where the Type I massive spinors become sub-Planckian . 

\paragraph{Type I Gauge Instanton.}  

Let us now briefly consider the gauge instanton. In \cite{witten_d-branes_1998}, it is argued that the \textit{perturbative} gauge group of the Type I vacuum actually corresponds to $\mO(32)$ instead of $\SO(32)$, being the structure group of a real Chan-Paton bundle. Via an analogous analysis to the case of the gauge spinor, the instanton is argued to be an object implementing the breaking of $\mO(32)$ to $\SO(32)$. On the heterotic side, the instanton action goes to zero, as the perturbative vaccum of the heterotic theory does not posess the disconnected $\B{Z}_2$ of $\mO(32)$ (or its $\Pin^-$-lift) as a symmetry. This, again, is consistent with the ECC picture that there be a regime in which the instanton action becomes small. 

\paragraph{} \hspace{2mm} In summary, we have seen that objects trivializing bordism classes carrying $\KO$-theoretic charges associated to Type I D-branes all find a limit in moduli where they become sub-Planckian. Several of these objects, including the torsion-valued gauge spinor and instanton, are non-BPS defects which only become light in the corresponding dual heterotic picture, providing a window into the applicability of our conjecture outside the realm of the BPS spectrum. In the following section, we will consider further non-BPS objects, this time from the perspective of the heterotic theory.

\subsection{Non-Supersymmetric Heterotic Branes in 10D} \label{ssec:non-susy-heterotic}

In this section, we study the brane solutions presented in \cite{kaidi_non-supersymmetric_2023, kaidi_non-supersymmetric_2024}, which are also non-BPS defects, this time visible from the \textit{heterotic} theory. In the heterotic picture, there are no D-branes or $\KO$-theoretic charges in sight, but one still has a gauge bundle with gauge group either $G = \Spin(32)/\B{Z}_2$ or $G = (E_8 \times E_8) \rtimes \B{Z}_2$. Associated to the topological gauge bundle configurations are various apparent bordism classes of $S^k$ equipped with a (homotopy class of a) map $S^k \to BG$; the trivialization of the class is a defect in dimension $(9-k)$. These defects are different from the Type I D-branes, as they correspond to \textit{trivializations} of the would-be bordism classes carrying gauge bundle configurations, while the Type I D-branes are best thought of as condensations the gauge bundles themselves. Accordingly, in the dual Type I frame (for the $\Spin(32)/\B{Z}_2$ theory) we should really view these branes as ``endpoints'' or ``junctions'' of D-branes. This motivates the consideration of at least some of these heterotic branes as \textit{relative} defects, as we will describe explicitly in what follows.

When first studying these branes, it appears that some of the brane solutions in \cite{kaidi_non-supersymmetric_2023, kaidi_non-supersymmetric_2024} appear to violate our conjecture! For example, the $k=8$ zerobrane in the $\Spin(32)/\B{Z}_2$ theory and the $k=4$ fourbranes in both theories\footnote{The $\Spin(32)/\B{Z}_2$ fourbrane is not discussed in the original references \cite{kaidi_nonsupersymmetric_2023, kaidi_non-supersymmetric_2024} as it is relative to a heterotic NS5 brane through the condition
\begin{equation} dH_3 = \frac{\a'}{2} \left( \tr R^2 - \tr F^2 \right). \end{equation}
In the context considered by the authors of \cite{kaidi_non-supersymmetric_2024, kaidi_non-supersymmetric_2023}, where an exact \textit{worldsheet} description in these brane backgrounds was desired, the limit $g_s \to 0$ was taken, sending the NS5 tension to infinity and confining all $\Spin(32)/\B{Z}_2$ fourbrane charges. In our setup at finite $g_s$ (and therefore finite $\a'$ relative to $m_{\m{pl}, 10}$), the $\Spin(32)/\B{Z}_2$ fourbrane becomes relevant.} naively appear to have \textit{infinite} tension per unit volume from an ADM computation. The reason for that is that the instanton configurations so defined only depend on the angular coordinates of the sphere, and a simple dimension counting argument shows that the kinetic term of the supergravity action is infrared divergent: Over $\B{R}^{k+1}$, the energy in a spherically-symmetric gauge field configuration up to an IR cutoff radius $L$ goes as 
\begin{equation}
    \label{eq:tension-het-brane}
     T \sim \int_0^L \d r \, r^{k} \, F \wedge * F.
\end{equation}
For profiles with $F \sim 1/r^2$ that have fixed instanton number, this diverges as $L^{k-3}$, which is in apparent tension with the ECC for a zerobrane (with $k = 8$) and a fourbrane (with $k = 4$).

In this section we point out the subtlety that addresses this problem, and shows that there is no contradiction after all. Moreover, this discussion helps elucidate how our discussion applies to low energy global symmetries that end up \emph{gauged} in the UV theory, as opposed to broken. Briefly, the branes with IR-divergent tension do not independently trivialize any bordism classes, and should indeed be viewed as relative defects attached to extended solitonic branes. We thus emphasize why bordism is the \textit{necessary} characterization of the charges of objects whose tensions are constrained by the ECC. 

We will explain in detail how this works for the particular case of the zerobrane in the $\Spin(32)/\B{Z}_2$ theory, which is associated to the homotopy group $\pi_8(B\Spin(32)/\B{Z}_2)$, after which we briefly discuss the fourbranes, which have an analogous divergent tension. Finally, we comment on the sixbrane and sevenbrane solutions proposed by \cite{kaidi_non-supersymmetric_2024, kaidi_non-supersymmetric_2023}, which have no divergent tension and represent genuine, non-relative defects.

\paragraph{$\Spin(32)/\B{Z}_2$ Zerobrane.} We explain in more detail the example of the zerobrane in the $\Spin(32)/\B{Z}_2$ theory. This object sources charge associated to a nontrivial homotopy class
\begin{equation} \pi_8(B\Spin(32)/\B{Z}_2) = \pi_7(\Spin(32)/\B{Z}_2) = \B{Z} \end{equation}
along a transverse $S^8$, and the characteristic class which detects this charge is $\m{ch}_{4}(E) = \tr_v F^4$, where $E$ a $\Spin(32)/\B{Z}_2$ gauge bundle over $S^8$, $F$ is the $\Spin(32)/\B{Z}_2$ field strength, and $\tr_v$ is the vector representation trace. There is, however, a non-trivial Bianchi identity for the seven-form $H_7 = \star H_{3}$ \cite{freed_dirac_2001}:
\begin{equation} \label{eq:dH7-bianchi-identity} dH_7  = \m{ch}_{4}(E) - \dfrac{1}{48}p_{1}(X)\m{ch}_{2}(E)+\dfrac{1}{64}p_{1}(X)^2-\dfrac{1}{48}p_{2}(X)    \end{equation}
The bordism picture for this system follows straightforwardly. Naively, one would associated to the zerobrane to a bordism class under the map
\begin{equation}\pi_7(\Spin(32)/\B{Z}_2) \longrightarrow \Omega_8^{\m{String}}(B \Spin(32)/\B{Z}_2)  \end{equation}
However, the Bianchi identity for $H_7$ enforces the consistency condition $\tr_v(F^4) = 0$ on all bordism classes of 8-manifolds for the theory.\footnote{The appropriate tangential structure for a theory which respects the Bianchi identity \eqref{eq:dH7-bianchi-identity} is \textit{twisted fivebrane structure} where the characteristic class $\tr_v F^4$ twists the background fivebrane structure $\frac{1}{6} p_2 = 0$. This is analogous to the Bianchi identity for $H_{3}$ representing a twisted String structure. The corresponding bordism group would be denoted schematically as $\Omega^{\m{Fivebrane}-\Spin(32)/\B{Z}_2}_8(\m{pt})$.} Accordingly, the would-be global symmetry associated to this bordism charge is \textit{gauged}, not broken, just like the example in Sec. \ref{ssec:gauged-bordism-defects}, and the proposed zerobrane does not independently trivialize a bordism class.

\begin{figure}[ht]
    \centering
    \includegraphics[width=0.6\linewidth]{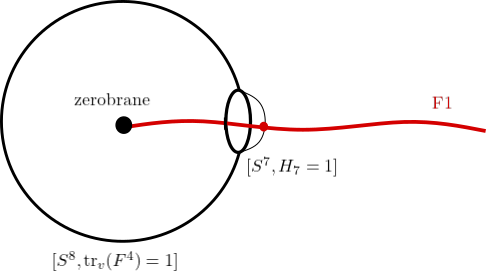}
    \caption{An illustration of the heterotic zerobrane as a relative defect, trivializing the singular manifold consisting of $S^8$ equipped with a gauge bundle with nontrivial $\tr_v F^4$ and an F1-string as a $H^7$-source.}
    \label{fig:zerobrane}
\end{figure}

To construct a zerobrane carrying charge associated to $S^8$ with one unit of $\tr_v F^4$, we must cut open an $S^{7}$ boundary and thread $1$ units of $H_7$ flux through it; this \emph{open} manifold is a consistent background for the string. In particular, it provides a bordism from the class $[S^7_{H_7}]$ of $S^{7}$ with $H_7$ flux to the trivial manifold; that nullbordism corresponds to the zerobrane. It is separately true, however, that the same class $[S^7_{H_7}]$ can be trivialized by the fundamental heterotic string sourcing the same flux. The joint trivialization of this singular 8-manifold in the category of bordisms \textit{with defects} yields precisely a open heterotic string ending on the zerobrane, following the discussion in Sec. \ref{ssec:gauged-bordism-defects}. A schematic depiction of this is given in Fig. \ref{fig:zerobrane}. 

Another way to think about it, going back to \cite{polchinski_open_2006}, is that the spherically symmetric gauge bundle configuration can itself collapse along a string, ``extending'' the solitonic string that ends on the zerobrane. The endpoint of this collapse is nothing other than an infinitely-extended long solitonic string in both directions. In fact, it was explicitly calculated in \cite{kaidi_non-supersymmetric_2024} that a (radially constant) spherically symmetric gauge configuration suffers a tachyonic instablity, rendering it unstable to collapse to a ``thick'' string. Further evidence lies in the presence of 24 chiral fermion zero modes along radial direction, flowing towards the core of the zerobrane, which will be discussed in upcoming work \cite{AlvarezGarcia-heterotic-202x}.

We can explicitly see the relation between this picture and the observed divergence in the zerobrane tension. One would expect the diverging tension due to a string to scale as $L$, where $L$ is the string length. However, as per Eq. \eqref{eq:tension-het-brane}, the spherical gauge bundle configuration described in \cite{kaidi_non-supersymmetric_2024} produces a divergence that goes as $L^5$. A dynamical collapse upon a thick ``flux tube'' string changes the divergence appropriately: Recall that $F \wedge \star F \sim r^{-4}$, so that
\begin{equation}T \sim \left( \displaystyle{\int_{0}^{L}  dr \, r^8 r^{-4} } \right) \bigg/ \left( {\displaystyle{\int_{0}^{L}  dr \, r^7 r^{-4} }} \right) \sim L,  \end{equation}
as expected. 

A loophole around the collapse of a spherical zerobrane to a long string was also observed in \cite{kaidi_non-supersymmetric_2024}, and it was argued that one could engineer stable configurations which \textit{fix} the boundary conditions at infinity as a spherically symmetric gauge configuration. However, the price we pay is that the gauge bundle now must vary in the \textit{radial} direction, deviating increasingly from the spherical profile as one approaches the core of the zerobrane. Thus, one expects that a non-supersymmetric attractor mechanism shifts the gauge profile toward the tension-minimizing thick string configuration, and there is no contradiction with the ECC.

We leverage this discussion to remark that this is a generic paradigm for low-energy global symmetries that end up being gauged in a UV theory: the new dynamical objects appear as configurations with boundaries, which in retrospect define a class not in SUGRA bordism, but in a bordism theory with defects.\footnote{We thank Jake McNamara for pointing this out to us.} Nonetheless, it still seems to be true that the tension of these composite objects is not parametrically larger than the Planck mass in some regime of moduli, as explained by this example.

\paragraph{Other Non-BPS Heterotic Branes.} The fourbranes with seemingly divergent tensions have a similar subtlety which is resolved in an analogous manner. For example, the fourbrane in the $(E_8 \times E_8) \rtimes \B{Z}_2$ theory is presented as a soliton with charge valued in $\mathbb{Z} \times \mathbb{Z}$. The anti-diagonal configuration with instanton charges $(n, -n)$ was in particular considered in \cite{kaidi_non-supersymmetric_2024, kaidi_non-supersymmetric_2023}, and it admits an M-theory lift as a pair of oppositely charged solitons living inside each Horava-Witten wall as the boundary of an M5-brane stretching across the interval \cite{bergshoeff_open_2006}. In analogy with the previous example, the apparent infrared-divergent tension can be understood as the contribution of this long membrane stretched across the M-theory interval. In the interior regime of the moduli space $g_s \simeq 1$, and the length $l$ of the M-theory interval is at the 11-dimensional Planck scale, the heterotic fourbrane has effective tension set by $T_{\m{M5}} l \simeq m_{\m{pl}, 11}^5$, consistent with the ECC. Note, that the $E_8 \times E_8$ fourbrane should be seen as an intrinsically 11-dimensional (relative) defect, at least for $g_s \simeq 1$.

There are two other branes presented in \cite{kaidi_non-supersymmetric_2024, kaidi_non-supersymmetric_2023} --- a sixbrane in the $\Spin(32)/\B{Z}_2$ theory carrying charge associated to $\pi_2(B\Spin(32)/\B{Z}_2) = \B{Z}_2$ and a sevenbrane in the $(E_8 \times E_8) \rtimes \B{Z}_2$ theory carrying charge associated to $\pi_1(B((E_8 \times E_8) \rtimes \B{Z}_2)) = \B{Z}_2$. These non-BPS torsion-charged defects suffer no infrared divergences \cite{kaidi_non-supersymmetric_2024}, and are thus genuine, non-relative defects. The ECC predicts, accordingly, that they carry Planckian tension in an appropriate regime of $g_s$. For the sixbrane, for which a supergravity solution is described in \cite{kaidi_non-supersymmetric_2024, kaidi_non-supersymmetric_2023}, we find that $e^\Phi \to \infty$ as $r$ approaches the horizon, so we predict that this defect should become sub-Planckian in the dual Type I regime. Note that this defect does \textit{not} represent a Type I D-brane --- as we discussed in Sec. \ref{sssec:type-i-dbranes}, the $\B{Z}_2$ it carries is not $\KO$-theoretic in origin.

\subsection{Torsion Cycles in Calabi-Yau Compactifications} \label{ssec:implications-geometry}

In this section, we outline non-trivial examples of branes wrapping torsion cycles in Calabi-Yau compactifications. Specifically, the ECC places constraints on the volumes of middle-dimensional cycles in a Calabi-Yau threefold in precisely such a way that the corresponding objects in the compactified theory (obtained by wrapped branes) have sub-Planckian tension. This is especially relevant in the case of \textit{torsional} middle-dimensional classes $H^n(X, \B{Z})_{\m{tors}}$ of a CY $n$-fold $X$ which admit no continuous moduli, as dimensional reduction over $X$ yields objects whose tensions (in lower-dimensional Planck units) are insensitive to the overall CY volume. Accordingly, for these objects the ECC specifically predicts that their volumes can never be ``too large'' (relative to the volume of the entire CY). 

Let us examine this in more detail as follows. For compactification on a CY $n$-fold $X$ (of real dimension $2n$) down to $d$ dimensions, the lower dimensional Planck mass depends on the volume of $X$, denoted $V_{X}$ and the $10d$ Planck mass $M_{pl}$ as

\begin{equation} M_{\text{pl}, d}^{d-2} = V_{X}  M_{\text{pl}}^8   \end{equation}

If a $p$-brane wraps a $p$-cycle $\Sigma \subset X$, the corresponding particle in the lower dimensional theory will have mass 
\begin{equation}m = V_{\Sigma} T , \end{equation}
where $T$ is the tension of the wrapping brane, possibly dependent on higher-dimensional moduli. For some special values of $(p,2n)$, the mass of the particle $m$ relative to the lower dimensional Planck mass does \emph{not} scale with the overall volume modulus of $X$. To find the appropriate values of $(p, 2n)$, note that $V_\Sigma \sim V_X^{\frac{p}{2n}}$, so 
\begin{equation}  \frac{m}{m_{\m{pl}, d}} \simeq V_\Sigma V_X^{-\frac{1}{d-2}} m_{\m{pl}, 10}^{-\frac{8}{d-2}}, \end{equation}
which is independent of $V_X$ provided that
\begin{equation} \label{eq:dimension-brane-cycle} \frac{p}{2n} = \frac{1}{d-2} = \frac{1}{8-2n}. \end{equation}
Going beyond just Calabi-Yau compactifications for a moment, there are three cases when this happens: real 3-cycles on complex threefolds for $(p, 2n) = (3, 6)$, real 1-cycles in complex surfaces for $(p, 2n) = (1, 4)$, and the top 7-cycle on a \textit{real} sevenfold for $(p, 2n) = (7, 7)$. 

In particular, if a middle-dimensional cycle on a Calabi-Yau threefold is a torsion class in homology, then it does not have a continuous volume modulus that would allow us to shrink it. Torsion cycles have zero periods and deformations thereof cannot be parametrized by a map to a period domain. In fact, any deformation of a torsion cycle would break all supersymmetry, and we expect such directions to have potential barriers obstructing an extended attractor flow (in the sense described in Sec. \ref{sec:applications-pheno}). Moreover, one cannot shrink or enlarge the entire CY to change the mass of the lower dimensional object --- these masses are insensitive to an overall volume rescaling of the whole $X$ by Eq. \eqref{eq:dimension-brane-cycle}. Our conjecture then puts a bound on the volume of this torsion cycles
\begin{equation} \label{eq:tension-bound-CY-cycle}
    T(\Phi_{0})V_{\Sigma} \lesssim M_{pl,d},
\end{equation}
where $\Phi_{0}$ is some appropriate attractor value of the higher-dimensional moduli. If the brane has an attractor solution that sends $\Phi_{0}$ to an infinite distance locus, than the above bound would be trivially satisfied. On the other hand, an object for which the attractor point is at the interior of moduli space will have a nonzero $\tau$ at the core and will put a nontrivial bound on $V_{\Sigma}$. For the case when a D3-brane wraps this cycle, the dilaton profile is constant and the tension does not depend on it, so we have a non-trivial bound on $V_{\Sigma}$. 

For an explicit example, consider the Calabi-Yau threefold $X = (\m{K3} \times T^2)/\B{Z}_2$, where $\B{Z}_2$ acts as a freely-acting involution of $\m{K3}$ and sends $x \mapsto -x$ on $T^2$. A calculation with the universal coefficient theorem yields $H_3(X, \B{Z})_{\m{tors}} = \B{Z}_2$, so there is a nontrivial torsion 3-cycle over $X$.  Accordingly, the ECC applied to D3-branes wrapping this cycle \textit{predicts} that the volume of the $\B{Z}_2$-torsion cycle of $X$ is of order 1. Recently, computational advances have made possible numerical calculations of CY metrics \cite{anderson_lectures_2023, larfors_numerical_2022, berglund_machine_2024}, and it would be interesting to compute torsion 3-cycle volumes in various examples as a test of the ECC. 

\section{Applications and Phenomenology} \label{sec:applications-pheno}

In this section, we wish to consider various applications in phenomenology of the ECC. The ECC is especially powerful when applied to non-supersymmetric objects, which could arise both in theories which are supersymmetric as well as in theories which have no supersymmetry. In the case of a theory that is supersymmetric, we are predicting that there is a point in the moduli space of the theory which gives a minimal mass/tension for the constituent. For example, consider a supersymmetric theory with $\mathbb{Z}_n$ gauge symmetry; we predict that there is a point on the moduli space where a charged particle generating $\mathbb{Z}_{n}$ has mass at most order 1 in Planck units. 

The predictions will be even more striking for non-supersymmetric theories, as is the case for our universe. In such theories we do not have massless modes, but for the case of positive cosmological constant, we may have light modes which play a similar role to the moduli fields in the supersymmetric case. For example in our universe we would be predicting that at some value of such light scalar fields, the magnetic monopole will have mass less than $m_{pl, 4}$ and not just $m_{pl,4}/\sqrt{\alpha}$ as predicted by the Weak Gravity Conjecture (WGC) applied to the magnetic force.

\paragraph{Magnetic Monopole.}
 If the magnetic monopole admits an effective description in terms of Standard Model degrees of freedom, then its coupling to light scalars would intermediate a fifth force; however, there are very tight bounds on the strength of such interactions. In terms of excursions in field space as driven by some attractor mechanism, this in turn means that the value of these scalars in our universe cannot be too far from the attractor value set by the core of the monopole --- otherwise, the required coupling to light scalars for a large attractor flow would correspond to a large fifth force interaction. Roughly speaking, the asymptotic value of the light scalar is expected to be close to the attractor value, and in that case the present conjecture predicts that the \emph{observed} mass of the monopole is sub-Planckian (in $4d$ Planck units). This is the expectation in many GUT models, but here we are getting at it from the ECC, a Swampland principle! Note that the ECC strengthens the magnetic WGC by only a factor of $1/\sqrt{\alpha} \simeq 10$, but it could have been much more striking if it had been the case that, for instance, $\alpha \sim 10^{-20}$ instead.

\paragraph{Field Spaces with Potential.}

In this section, we make a brief comment about the case where a potential $V(\Phi)$ lifts the scalars off the massless moduli space $\mathcal{M}$. The dynamical minimization analogous to attractor flow then has to take into account the energy cost associated to this potential, and is effectively a trade-off between lowering the core tension of the object at the expense of driving scalars uphill the potential at an energy cost. In particular, the notion of geodesic distance appropriate for massless moduli is replaced by action-minimizing trajectories. Qualitatively, regions in the interior of field space with large potential will repel such trajectories, while those with large negative potential will focus them. Typically, if there are field space directions available, attractor mechanism will have the physical running solution dodge these large potential regions, and considerations about (almost) massless moduli should hold.  The ECC predicts that that these high potential regions will not be high enough as to spoil the sub-Planckian tension of generator defects, and this is in fact compatible with other Swampland principles. For example, if the attractor fixed point lies at infinite distance in field space, the Distance Conjecture implies that the potential should also decrease exponentially for the EFT of light modes to be valid -- in this regime there cannot be large potential hills at all.

\section{Discussion} \label{sec:discussion}

In this work, we have focused on the \textit{tensions} of elementary objects carrying bordism charge in quantum gravity. As a sort of dynamical extension of the Cobordism Conjecture, we conjectured that the tension of every generator of some bordism class is less than $m_{\m{pl}}$ in some region in moduli space, and moreover, the attractor mechanism drives the fields towards that regime as one approaches the core of the object. A diverse array of supersymmetric and non-supersymmetric examples illustrates several subtle ways in which our conjecture is observed to hold. We emphasize also the importance of regarding the \textit{charge} carried by an object as being detected by bordism --- in some examples (notably the heterotic zerobrane and fourbrane), surprising aspects of the dynamics are explained only by the bordism definition of charge. 

Our conjecture is motivated by the following existing conjectures of the Swampland program:
\begin{itemize}
    \item The Cobordism Conjecture predicts the existence of objects carrying ``universal'' bordism charge, but does not stipulate anything about the dynamics of these objects.
    \item The Weak Gravity Conjecture constrains the tension of charged objects, but only in the case of $U(1)$ gauge charges.
    \item The Distance Conjecture predicts the dynamics of objects near infinite distance in the moduli space of vacua, but does not constrain the geometry or the spectrum elsewhere.
\end{itemize}
We argue for the existence of \textit{regions of moduli space} (possibly containing the interior) where generators of \textit{bordism charge} have \textit{constrained tension} less than $m_{\m{pl}}$. We emphasize also that the ECC \textit{should not be surprising} --- it is in some sense the most obvious intersection of the three conjectures above. 

It would be most interesting to find more non-supersymmetric examples to check the ECC, or use the ECC to make novel predictions for these cases. It would also be interesting to find more phenomenological applications of ECC; in particular, a better understanding of the non-supersymmetric examples in which we expect to have potentials would need to be developed. We leave these works for the future.

\subsubsection*{Acknowledgments} 

This work is supported in part by a grant from the Simons Foundation (602883, CV) and the DellaPietra Foundation. We would like to thank Rafael \'{A}lvarez-Garc\'{i}a, Matilda Delgado, Muldrow Etheredge, Max H\"{u}bner, Ashoke Sen, Ethan Torres, and Kai Xu for stimulating discussions; as well as Jake McNamara and Miguel Montero for discussions and comments on the draft. We would also like to thank the Simons Summer Physics Workshop 2025 for providing a productive research environment where this project was initiated.

\appendix

\section{Compactifications with Maximal Supersymmetry} \label{ssec:compactifications-maximal-susy}

In this appendix, we study a more general class of examples of the ECC arising from compactifications preserving 32 supercharges. These examples have several moduli, and the associated objects in the $\frac{1}{2}$-BPS spectrum have moduli-dependent tensions. We wish to see that there is a regime in moduli at which \textit{all} of these tensions are Planckian. In the case without axionic moduli, we will see that this regime is precisely the vicinity of the ``self-duality'' point at which all of the radii of the compact $T^k$ are Planckian (in higher-dimensional Planck units). Though these examples have been well-understood for some time, they provide an explicit construction of the conjectured ``regime of elementarity'' for the (generators of the) entire $\frac{1}{2}$-BPS spectrum, which provides an alternate characterization of the so-called \textit{desert point} of \cite{van_de_heisteeg_moduli-dependent_2022, van_de_heisteeg_species_2023}. 

Consider M-theory compactified on $T^k$. The moduli space of maximal supergravity in $d = (11-k)$ dimensions is given by 
\begin{equation} \C{M}_d = E_{k(k)}(\B{Z}) \backslash E_{k(k)} / K_k, \end{equation}
where $K_k \subset E_{k(k)}$ is the maximal compact subgroup of the split real form $E_{k(k)}$ of the Lie group $E_k$ and $E_{k(k)}(\B{Z})$ is the discrete \textit{duality group}, which is expected to be a (spontaneously broken) discrete gauge symmetry of the theory. At the level of the low-energy effective action, the action of the global symmetry $E_{k(k)}(\B{R})$ on the scalars parametrizing $\C{M}$ is manifest; this symmetry is broken generically by the inclusion of dynamical charged objects in the massive spectrum of the theory. These observables in turn carry representations of the unbroken gauge symmetry $E_{k(k)}(\B{Z})$, and their tensions are expressible as functions on the marked moduli space $\hat{\C{M}} = E_{k(k)}/K_k$ \cite{raman_swampland_2024}. We will see that all of the $\frac{1}{2}$-BPS tensions go to zero in appropriate limit of the moduli; moreover, there is a regime in the interior of the saxionic (noncompact) moduli space where \textit{all} of the tensions are Planckian (up to $O(1)$ factors). 

We now review the construction of the $\frac{1}{2}$-BPS spectrum of maximal supergravity in \cite{obers_u-duality_1999, iqbal_mysterious_2001} in terms of duality representations, explicitly writing down the tensions of $\frac{1}{2}$-BPS objects as functions of the saxionic moduli (namely, the radii of the internal torus) which take values in the split Cartan of $E_{k(k)}$. To be precise, let us restrict ourselves to the locus $\C{M}_0 \subset \C{M}$ of the moduli space $\C{M} = E_{k(k)}(\B{Z}) \backslash E_{k(k)} / K_d$ corresponding to \textit{rectangular} tori with zero axionic moduli: 
\begin{equation} g_{ij} = R_i^2 \delta_{ij}, \q C_{ijk} = 0. \end{equation}
The locus $\C{M}_0$ admits a covering $\hat{\C{M}}_0 \into \hat{\C{M}}$ given by the (identity component of the) split Cartan $H_d \subset E_{k(k)}$, where $H_d \simeq \B{R}_{> 0}^d$. The subgroup of $E_{k(k)}(\B{Z})$ preserving $\hat{\C{M}}_0 = H_d \simeq \B{R}^d_{> 0}$ is precisely the Weyl group $W(E_d)$ of $E_d$. In total, 
\begin{equation} \C{M}_0 = H_d/W(E_d), \end{equation}
and it is this space we study more closely. 

In more detail, recall that the full duality group $E_{k(k)}(\B{Z})$ is generated by two different types of elements -- generators of `Weyl' type and of `Borel' type. The `Weyl' generators of $E_{k(k)}(\B{Z})$ are orientation-preserving elements of $E_{k(k)}(\B{Z})$ generating the Weyl group $W(E_d) \subset E_{k(k)}(\B{Z})$ which acts on $\hat{\C{M}}_0$, and they (roughly) act by T-duality on three independent radii (the action of a single T-duality is an orientation-reversing element of the full bosonic duality group). 

To construct the Weyl generators of $E_{k(k)}(\B{Z})$, let us first turn our attention to the perturbative string theory case, where we have access to T-duality as a known symmetry of the theory. In perturbative Type II string theory compactified on $T^{d-1}$ (hence M-theory on $T^d$), we define a $T$-transformation $T_{ij} \in \SO(d-1, d-1; \B{Z}) \subset E_{k(k)}(\B{Z})$ acting on $d$ internal compact radii $R_1, \cdots, R_{d-1}$ via two T-dualities on $R_i, R_j$ followed by an exchange $R_i \leftrightarrow R_j$:
\begin{equation} T_{ij} : (R_i, R_j, g_s) \longmapsto \left( \frac{l_s^2}{R_j}, \frac{l_s^2}{R_i}, \frac{g_s l_s^3}{R_i R_j} \right). \end{equation}
Here we have set $l_s$ to be the string length. Defining further $l_{\m{pl}, 11}$ to be the 11-dimensional Planck length, we may then relate $g_s$ to $l_s$ and the length of the M-theory circle $R_0$ via 
\begin{equation} R_0 = g_s^{2/3} l_{\m{pl}, 11}, \q l_{\m{pl}, 11}^3 = g_s l_s^3.  \end{equation}
Correspondingly, the T-duality `Weyl' generator in M-theoretic variables is given by 
\begin{equation}  T_{0ij} \equiv T_{ij} : (R_i, R_j, R_0, l_{\m{pl}, 11}^3) \longmapsto \left( \frac{l_{\m{pl}, 11}^3}{R_j R_0}, \frac{l_{\m{pl}, 11}^3}{R_i R_0}, \frac{l_{\m{pl}, 11}^3}{R_i R_j}, \frac{l_{\m{pl}, 11}^6}{R_i R_j R_k}\right) . \end{equation}
The action of $T_{0ij}$ is completely symmetric in $R_0, R_i, R_j$; furthermore, since there is a manifest $S_n$ symmetry exchanging the radii $R_0, \cdots, R_{d-1}$, we expect the correspondingly-defined generators $T_{ijk}$ to also be `Weyl'-type symmetries in $E_{k(k)}(\B{Z})$, which generate the Weyl group $W(E_d)$. Indeed, the Weyl group of $E_{d}$ can be expressed as
\begin{equation} W(E_d) = S_d \bowtie \B{Z}_2, \end{equation}
where the $\bowtie$ denotes a \textit{knit product} of two non-commuting subgroups, $S_d$ acts via permutation on $\{R_0, \cdots, R_{d-1}\}$, and $\B{Z}_2$ acts via the T-transformation $T_{ijk}$ as given above. 

Although they are orthogonal to our discussion, we briefly comment on the nature of the `Borel'-type generators of $E_{k(k)}(\B{Z})$. These correspond roughly to axionic monodromy matrices, which generically generate Borel-type nilpotent subgroups of the duality group. For instance, the modulus $C_{ijk}$ associated to the M-theory $C_3$-field wrapping three compact legs enjoys a periodicity $C_{ijk} \mapsto C_{ijk} + 1$ associated to the $U(1)$-valued (higher) connection in the 11-dimensional theory. Generators of this form enlarge $W(E_d)$ to the full duality group $E_{k(k)}(\B{Z})$. In what follows, we will restrict our attention to the representations of $W(E_d)$, which comprise the interesting ``nonperturbative'' duality transformations. 

Let us define the ``dual weight space'' $\f{h}$ of $E_d$, canonically isomorphic to the Cartan subalgebra of $E_d$, such that $\f{h} \simeq \B{R}^d$ as vector spaces. We parametrize $\f{h}$ by the coordinates $\phi_i = \log 2\pi R_i$ for $i = 0, \cdots, d-1$, and we also define $\phi_{-1} = 3 \log l_{\m{pl}, 11}$. In $(11-d)$-dimensional Planck units, the condition that $l_{\m{pl}, 11-d}^{9-d} = l_{\m{pl}, 11}^9 ((2\pi R_0) \cdots (2\pi R_{d-1}))^{-1}$ yields
\begin{equation} \phi_{-1} = -\frac{1}{3} \sum_{i = 0}^{d-1} \phi_i  \end{equation}
so $\phi_{-1}$ is not independent from the $\phi_0, \cdots, \phi_{d-1}$. We nevertheless sometimes retain an explicit dependence on $\phi_{-1}$ in what follows to make contact with the notation in \cite{obers_u-duality_1999, iqbal_mysterious_2001}, and we denote the total space parametrized by $\phi_{-1}, \phi_0, \cdots, \phi_{d-1}$ as $\tilde{\f{h}}$, which we will call the ``extended dual weight space''. The Weyl generators $S_\s$ for $\s \in S_d$ and $T_{ijk}$ then act canonically as a Coxeter group on $\tilde{\f{h}}$ as parametrized by the $\phi_i$. The group $W(E_d)$, equipped with its action on $\tilde{\f{h}}$, preserves the following indefinite metric
\begin{equation} ds^2 = -\d\phi_{-1}^2 + \sum_{i = 0}^{d-1} \d\phi_i^2. \end{equation}
Moreover, the Weyl group preserves the condition $3\phi_{-1} + \phi_0 + \cdots + \phi_{d-1} = 0$, restricting therefore to an action by Coxeter reflections on the subspace $\f{h}$ equipped with a positive-definite metric. The corresponding dual action on $\f{h}^\vee$ is shown in \cite{obers_u-duality_1999} to precisely reproduce the action of $W(E_8)$ on the weight space of $E_8$, justifying our calling $\f{h}$ the ``dual weight space''. 

Associated to every positive weight $\l \in \f{h}^\vee$, there is a corresponding representation of $E_{k(k)}(\B{Z})$, which therefore admits an action of $W(E_d)$ that permutes its constituent weight spaces. (The action of the `Borel' generators on $\l$ shift $\l$ by integer multiples of roots of $E_d$.) Accordingly, we find that each state in the effective theory parametrized by $\hat{\C{M}}$ (and therefore by $\hat{\C{M}}_0 \subset \hat{\C{M}}$) necessarily carries an action of $W(E_d)$ and therefore assembles in a multiplet of weight spaces $V_\l \subset \f{h}^\vee$ which are exchanged by the action of $W(E_d)$ (plus additional Borel generators). The condition that there be \textit{finitely} many weight spaces in each such multiplet then restricts our attention to positive, \textit{dominant integral} weights $\l \in V$, attached to which there is a \textit{finite-dimensional} irreducible representation $L_\l$ of $E_{d}(\B{C})$ that decomposes as the sum of finitely many weight spaces $V_\l$. These weight spaces carry an action of $W(E_d)$, the orbits of which correspond to subspaces of $L_\l$ spanned by $V_\rho$ at fixed $\abs{\rho}$. 

To summarize, we have seen that each object parametrized by $\hat{\C{M}}_0$ is necessarily associated to a finite direct sum of weight spaces $V_\rho$ with an action of $W(E_d)$ associated to an irrep $L_\l$ with positive dominant integral $\l$. It was the work of \cite{iqbal_mysterious_2001} that showed that the contents of the $\frac{1}{2}$-BPS spectrum of M-theory on $T^d$ are realized precisely by multiplets associated to `minuscule' weights $\l \in \f{h}^\vee$ generating fundamental representations $L_\l$ of $E_d(\B{C})$. Even more remarkably, it was shown that there is an $W(E_d)$-equivariant equivalence of the extended dual weight space $\tilde{\f{h}}$ with $H^2(\B{B}_d, \B{R})$, where $\B{B}_d$ is the $d$th del Pezzo surface. The $\frac{1}{2}$-BPS spectrum is then characterized by minuscule weights in the (extended) weight lattice $ H_2(\B{B}_d, \B{Z}) \subset U $, with the tension of $\C{C} \in H_2(\B{B}_d, \B{Z})$ given by the (exponentiated) volume of $\C{C}$:
\begin{equation} T_{\C{C}} = 2\pi \exp(\omega(\C{C})). \end{equation}
for a choice of Kähler class $\omega \in H^2(\B{B}_d, \B{R})$. This identification was referred to as ``mysterious duality'' in \cite{iqbal_mysterious_2001}. More explicitly, for a weight vector $\l \in U$ expressed as $l = (n, m_0, \cdots, m_{d-1})$ and a Kähler class $\w \in U^\vee$ with coordinates $\w = (3\log l_{\m{pl}, 11}, \log 2\pi R_0, \cdots, \log 2\pi R_{d-1})$ corresponding to a choice of a point in $\hat{\C{M}}_0$, 
\begin{equation} T_\l = 2\pi e^{\ev{\l, \w}} = 2\pi l_{\m{pl}, 11}^{-3n} (2\pi R_0)^{m_0} \cdots (2\pi R_{d-1})^{m_{d-1}}. \end{equation}

Thus far, the entire discussion has been a review of \cite{obers_u-duality_1999, iqbal_mysterious_2001}. It is now a simple observation to understand precisely how the $\frac{1}{2}$-BPS tensions of vary with the moduli. In particular, for a choice of $\frac{1}{2}$-BPS state corresponding to a highest-weight vector $\l \in U$, we compute the tensions of the duality orbit associated to $\l$ by taking $\ev{\w, \rho}$ for $\rho \in W(E_d)(\l)$. There is a (non-canonical) isomorphism $\f{h}^\vee \simeq \f{h}$, and accordingly, for each $\rho$, we can choose the value of the moduli $\omega$ oriented along $-\rho$. Then we find immediately that the tension $T_\rho$ is given by 
\begin{equation} T_\rho \sim e^{-\abs{\rho}^2}, \end{equation}
which can be made arbitrarily small in an appropriate regime of the moduli space. Furthermore, at the point $\w = 0$, which is preserved by the entire group $W(E_d)$, we find that 
\begin{equation} T_\rho = 2\pi, \q \rho \in \tilde{\f{h}}^\vee\end{equation}
Thus, it is precisely at the self-dual point, at least in $\hat{\C{M}}_0 = H_0 = \exp(\f{h})$, at which the tension of \textit{every} $\frac{1}{2}$-BPS object is exactly Planckian. We expect that this analysis holds true for other examples of self-duality/enhanced discrete gauge symmetry loci in the interiors of moduli spaces.

\section{Closest Approach and Stretched Horizons} \label{app:stretched-horizon-entropy}

In this Appendix, we provide a candidate proposal for the near-horizon regime of singular-horizon ``dynamical cobordism'' solutions, with a focus on the example of D$p$-branes for $p \neq 3$. In particular, in Sec. \ref{ssec:examples-d-branes}, we showed that D$p$-branes for $p \neq 3$ have moduli-dependent tensions that, in an appropriate limit of $g_s$, become sub-Planckian. We then argued in Sec. \ref{sec:attractor} that this limit is engineered in the supergravity solution by the attractor mechanism. However, it remains unclear as to exactly \textit{what} ``core tension'' should be measured after taking the energy stored in the scalar field gradients imposed by the attractor flow equations. 

For the case of D$p$-branes with $p \neq 3$, the associated supergravity solutions have \textit{singular horizons} in the sense of \cite{bedroya_holography_2025}. Accordingly, the horizon does not lie at the end of an infinite throat, and we immediately run into a problem. The effective supergravity theory has UV cutoff set by the species scale $\Lambda_{\m{sp}}$, and we expect that we cannot sensibly describe features of the theory at scales beyond $\Lambda_{\m{sp}}$. As a result, we cannot approach within a distance less than $\ell_{\m{sp}} = 1/\Lambda_{\m{sp}}$ of the horizon before the EFT breaks down. 

Actually, the picture is substantially more complicated, since the species scale is properly understood as the cutoff scale associated to EFT backgrounds with constant background moduli values \cite{van_de_heisteeg_moduli-dependent_2022, van_de_heisteeg_species_2023}. Thus, in our running D$p$-brane solutions, the cutoff may deviate significantly from $\Lambda_{\m{sp}}(\phi_0)$ at the asymptotic values of moduli $\phi_0$. Nevertheless, we always have the bound $\Lambda_{\m{sp}} \leq m_{\m{pl}, d}$ for a $d$-dimensional EFT. Thus, at a coarsest approximation, we recognize that we may not approach the core of the brane at a distance less than $l_{\m{pl}, 10}$, which we term the ``distance of closest approach''. (This locus has also been variously called a \textit{stretched horizon}, cf. \cite{susskind_stretched_1993})

We now study the attractor solutions for single D$p$-branes up to this closest-approach point. To begin, recall that the metric for a stack of extremal black D$p$-branes in the Einstein frame is given by 
\begin{equation} ds^2 = H(r)^{(p-7)/8} (-\d t^2 + \d \vec{x}^2) + H(r)^{(p+1)/8}(\d r^2 + r^2 \d \Omega_{8-p}^2). \end{equation}
with
\begin{equation} H(r) = 1 +  \frac{\a}{r^{7-p}}, \q \a = C_p l_{\m{p}, 10}^{7-p} N g_{s, \infty}^{-(3-p)/4}. \end{equation}
Here, $g_{s, \infty}$ is the value of the string coupling at asymptotic infinity and $N$ is the number of D$p$-branes in the stack. For our purposes, we are interested in the tensions of \textit{elementary} defects, so we take $N = 1$. 

When the geodesic distance from the core of the brane equals $l_{\m{pl}, d}$, one cannot trust the supergravity description any further. The geodesic distance from the core of the brane is computed in terms of the radial coordinate $r$ as follows: 
\begin{equation} d(r) = \int_0^r \d r' \, H(r')^{(p+1)/16}. \end{equation}
It is crucial to point out that by considering this geodesic distance, we intend to identify the \emph{submanifold} corresponding to the threshold of validity of the EFT description. The coordinate description of this submanifold is not a well-defined notion, but the submanifold itself certainly is. In the near-horizon limit for which $\a/r^{7-p} \gg 1$, we find that
\begin{equation} d(r) = \a^{(p+1)/16} r^{(3-p)^2/16} \simeq l_{\m{p}, 10} g_{s, \infty}^{-\frac{(p+1)(3-p)}{64}} \pfc{r}{l_{\m{p}, 10}}^{(3-p)^2/16} . \end{equation}
We compute also the value of $g_s$ as a function of the radial coordinate $r$: 
\begin{equation} g_s(r) = g_{s, \infty} H(r)^{(3-p)/4}. \end{equation}
For $p < 3$, the attractor mechanism sends $g_s(r) \to \infty$ as $r \to 0$, while for $p > 3$, $g_s(r) \to 0$ as $r \to 0$, just as described in Sec. \ref{ssec:examples-d-branes}. Thus, for a well-defined supergravity solution far from the horizon for a single D$p$-brane, we wish to take $g_{s, \infty} \gg 1$ for $p < 3$ and $g_{s, \infty} \ll 1$ for $p > 3$. These regimes correspond precisely to the ``worst-case'' limits of asymptotic moduli where the supergravity solution has asymptotic tension far greater than unity in Planck units. 

As we have seen, even in these worst-case regimes, the attractor flow drives the value of $g_s(r)$ towards the opposite limit as one approaches the singular horizon. However, we have also seen that we cannot go past the closest-approach point. The question therefore remains -- does the tension become sub-Planckian even at this point? We consider the closest-approach radius $r_*$ for which $d(r_*) = l_{\m{pl}, 10}$, at which point we require
\begin{equation} r_* \simeq l_{\m{pl}, 10} g_{s, \infty}^{\frac{p+1}{4(3-p)}}. \end{equation}
We note that
\begin{equation} \frac{\a}{r_*^{7-p}} \simeq g_{s, \infty}^{-(3-p)/4} g_{s, \infty}^{-\frac{(p+1)(7-p)}{4(3-p)}} = g_{s, \infty}^{-\frac{4}{3-p}}, \end{equation}
so that for either $p < 3$ and $g_{s, \infty} \lesssim 1$ or $p > 3$ and $g_{s, \infty} \gtrsim 1$, this quantity becomes much larger than unity, and the near-horizon approximation $H(r) \simeq \a/r^{7-p}$ is valid. We therefore calculate that 
\begin{equation} g_s(r_*) \simeq g_{s, \infty} \pfc{\a}{r_*^{7-p}}^{(3-p)/4} \simeq 1  \end{equation}
at the closest-approach point, and for this value of moduli, the tension of the D$p$-brane is exactly Planckian, as required by the ECC. Note that even if we were to use the naive species scale as a function of the running moduli, one would have $l_{\m{sp}}(r_*) = l_{\m{sp}}(g_s = 1) = l_{\m{pl}, 10}$, which is consistent with our \textit{ansatz} that the closest-approach distance be $l_{\m{pl}, 10}$ to begin with. This simple calculation therefore demonstrates that one does not actually need to trace the attractor flow beyond the naive validity of the EFT to reach a regime of moduli where the tension of an object is ECC-consistent.

\bibliographystyle{JHEP}

\bibliography{bib}

\providecommand{\href}[2]{#2}\begingroup\raggedright\begin{thebibliography}{10}

\bibitem{mcnamara_cobordism_2019}
J.~McNamara and C.~Vafa, \emph{Cobordism {Classes} and the {Swampland}},  Oct., 2019.
\newblock 10.48550/arXiv.1909.10355.

\bibitem{debray_chronicles_2023}
A.~Debray, M.~Dierigl, J.J.~Heckman and M.~Montero, \emph{The {Chronicles} of {IIBordia}: {Dualities}, {Bordisms}, and the {Swampland}}, .

\bibitem{montero_cobordism_2021}
M.~Montero and C.~Vafa, \emph{Cobordism {Conjecture}, {Anomalies}, and the {String} {Lamppost} {Principle}}, \href{https://doi.org/10.1007/JHEP01(2021)063}{\emph{JHEP} {\bfseries 01} (2021) 063}.

\bibitem{mcnamara_gravitational_2021}
J.~McNamara, \emph{Gravitational {Solitons} and {Completeness}},  Aug., 2021.
\newblock 10.48550/arXiv.2108.02228.

\bibitem{dierigl_swampland_2021}
M.~Dierigl and J.J.~Heckman, \emph{On the {Swampland} {Cobordism} {Conjecture} and {Non}-{Abelian} {Duality} {Groups}}, \href{https://doi.org/10.1103/PhysRevD.103.066006}{\emph{Physical Review D} {\bfseries 103} (2021) 066006}.

\bibitem{buratti_dynamical_2021}
G.~Buratti, M.~Delgado and A.M.~Uranga, \emph{Dynamical {Tadpoles}, {Stringy} {Cobordism}, and the {SM} from {Spontaneous} {Compactification}}, \href{https://doi.org/10.1007/JHEP06(2021)170}{\emph{Journal of High Energy Physics} {\bfseries 2021} (2021) 170}.

\bibitem{van_beest_lectures_2022}
M.~van Beest, J.~Calderón-Infante, D.~Mirfendereski and I.~Valenzuela, \emph{Lectures on the {Swampland} {Program} in {String} {Compactifications}}, \href{https://doi.org/10.1016/j.physrep.2022.09.002}{\emph{Physics Reports} {\bfseries 989} (2022) 1}.

\bibitem{agmon_lectures_2023}
N.B.~Agmon, A.~Bedroya, M.J.~Kang and C.~Vafa, \emph{Lectures on the string landscape and the {Swampland}},  Mar., 2023.
\newblock 10.48550/arXiv.2212.06187.

\bibitem{banks_symmetries_2011}
T.~Banks and N.~Seiberg, \emph{Symmetries and {Strings} in {Field} {Theory} and {Gravity}}, \href{https://doi.org/10.1103/PhysRevD.83.084019}{\emph{Physical Review D} {\bfseries 83} (2011) 084019}.

\bibitem{harlow_symmetries_2019}
D.~Harlow and H.~Ooguri, \emph{Symmetries in quantum field theory and quantum gravity},  June, 2019.
\newblock 10.48550/arXiv.1810.05338.

\bibitem{rudelius_topological_2020}
T.~Rudelius and S.-H.~Shao, \emph{Topological {Operators} and {Completeness} of {Spectrum} in {Discrete} {Gauge} {Theories}}, \href{https://doi.org/10.1007/JHEP12(2020)172}{\emph{Journal of High Energy Physics} {\bfseries 2020} (2020) 172}.

\bibitem{heidenreich_non-invertible_2021}
B.~Heidenreich, J.~McNamara, M.~Montero, M.~Reece, T.~Rudelius and I.~Valenzuela, \emph{Non-{Invertible} {Global} {Symmetries} and {Completeness} of the {Spectrum}}, \href{https://doi.org/10.1007/JHEP09(2021)203}{\emph{Journal of High Energy Physics} {\bfseries 2021} (2021) 203}.

\bibitem{arkani-hamed_string_2007}
N.~Arkani-Hamed, L.~Motl, A.~Nicolis and C.~Vafa, \emph{The {String} {Landscape}, {Black} {Holes} and {Gravity} as the {Weakest} {Force}}, \href{https://doi.org/10.1088/1126-6708/2007/06/060}{\emph{Journal of High Energy Physics} {\bfseries 2007} (2007) 060}.

\bibitem{harlow_weak_2023}
D.~Harlow, B.~Heidenreich, M.~Reece and T.~Rudelius, \emph{The {Weak} {Gravity} {Conjecture}: {A} {Review}}, \href{https://doi.org/10.1103/RevModPhys.95.035003}{\emph{Reviews of Modern Physics} {\bfseries 95} (2023) 035003}.

\bibitem{kaidi_non-supersymmetric_2024}
J.~Kaidi, Y.~Tachikawa and K.~Yonekura, \emph{On non-supersymmetric heterotic branes},  Nov., 2024.
\newblock 10.48550/arXiv.2411.04344.

\bibitem{kaidi_non-supersymmetric_2023}
J.~Kaidi, K.~Ohmori, Y.~Tachikawa and K.~Yonekura, \emph{Non-supersymmetric heterotic branes},  Mar., 2023.

\bibitem{delgado_finiteness_2024}
M.~Delgado, D.v.d.~Heisteeg, S.~Raman, E.~Torres, C.~Vafa and K.~Xu, \emph{Finiteness and the {Emergence} of {Dualities}},  Dec., 2024.
\newblock 10.48550/arXiv.2412.03640.

\bibitem{heckman_fate_2024}
J.J.~Heckman, J.~McNamara, M.~Montero, A.~Sharon, C.~Vafa and I.~Valenzuela, \emph{On the {Fate} of {Stringy} {Non}-{Invertible} {Symmetries}},  Jan., 2024.

\bibitem{sati_flux_2025}
H.~Sati and U.~Schreiber, \emph{Flux {Quantization}},  in \emph{Encyclopedia of {Mathematical} {Physics}}, pp.~281--324 (2025), \href{https://doi.org/10.1016/B978-0-323-95703-8.00078-1}{DOI}.

\bibitem{buratti_dynamical_2021-1}
G.~Buratti, J.~Calderón-Infante, M.~Delgado and A.M.~Uranga, \emph{Dynamical {Cobordism} and {Swampland} {Distance} {Conjectures}},  July, 2021.
\newblock 10.48550/arXiv.2107.09098.

\bibitem{ferrara_n2_1995}
S.~Ferrara, R.~Kallosh and A.~Strominger, \emph{N=2 {Extremal} {Black} {Holes}}, \href{https://doi.org/10.1103/PhysRevD.52.R5412}{\emph{Physical Review D} {\bfseries 52} (1995) R5412}.

\bibitem{moore_attractors_2003}
G.~Moore, \emph{Attractors and {Arithmetic}},  July, 2003.
\newblock 10.48550/arXiv.hep-th/9807056.

\bibitem{ferrara_black_1997}
S.~Ferrara, G.W.~Gibbons and R.~Kallosh, \emph{Black {Holes} and {Critical} {Points} in {Moduli} {Space}}, \href{https://doi.org/10.1016/S0550-3213(97)00324-6}{\emph{Nuclear Physics B} {\bfseries 500} (1997) 75}.

\bibitem{behrndt_classical_1997}
K.~Behrndt, G.L.~Cardoso, B.d.~Wit, R.~Kallosh, D.~Lüst and T.~Mohaupt, \emph{Classical and quantum {N}=2 supersymmetric black holes}, \href{https://doi.org/10.1016/S0550-3213(97)00028-X}{\emph{Nuclear Physics B} {\bfseries 488} (1997) 236}.

\bibitem{ferrara_supersymmetry_1996}
S.~Ferrara and R.~Kallosh, \emph{Supersymmetry and {Attractors}}, \href{https://doi.org/10.1103/PhysRevD.54.1514}{\emph{Physical Review D} {\bfseries 54} (1996) 1514}.

\bibitem{ruehle_attractors_2024}
F.~Ruehle and B.~Sung, \emph{Attractors, {Geodesics}, and the {Geometry} of {Moduli} {Spaces}},  Aug., 2024.
\newblock 10.48550/arXiv.2408.00830.

\bibitem{sen_black_2008}
A.~Sen, \emph{Black {Hole} {Entropy} {Function}, {Attractors} and {Precision} {Counting} of {Microstates}}, \href{https://doi.org/10.1007/s10714-008-0626-4}{\emph{General Relativity and Gravitation} {\bfseries 40} (2008) 2249}.

\bibitem{greene_stringy_1990}
B.R.~Greene, A.D.~Shapere, C.~Vafa and S.-T.~Yau, \emph{Stringy {Cosmic} {Strings} and {Noncompact} {Calabi}-{Yau} {Manifolds}}, \href{https://doi.org/10.1016/0550-3213(90)90248-C}{\emph{Nucl. Phys. B} {\bfseries 337} (1990) 1}.

\bibitem{obers_u-duality_1999}
N.A.~Obers and B.~Pioline, \emph{U-duality and {M}-{Theory}}, \href{https://doi.org/10.1016/S0370-1573(99)00004-6}{\emph{Physics Reports} {\bfseries 318} (1999) 113}.

\bibitem{iqbal_mysterious_2001}
A.~Iqbal, A.~Neitzke and C.~Vafa, \emph{A {Mysterious} {Duality}},  Dec., 2001.
\newblock 10.48550/arXiv.hep-th/0111068.

\bibitem{tachikawa_why_2019}
Y.~Tachikawa and K.~Yonekura, \emph{Why are fractional charges of orientifolds compatible with {Dirac} quantization?}, \href{https://doi.org/10.21468/SciPostPhys.7.5.058}{\emph{SciPost Physics} {\bfseries 7} (2019) 058}.

\bibitem{bergshoeff_iib_2007}
E.~Bergshoeff, J.~Hartong, T.~Ortín and D.~Roest, \emph{{IIB} seven–branes revisited}, \href{https://doi.org/10.1088/1742-6596/66/1/012054}{\emph{Journal of Physics: Conference Series} {\bfseries 66} (2007) 012054}.

\bibitem{brav_thin_2014}
C.~Brav and H.~Thomas, \emph{Thin monodromy in {Sp}(4)}, \href{https://doi.org/10.1112/S0010437X13007550}{\emph{Compositio Mathematica} {\bfseries 150} (2014) 333}.

\bibitem{heckman_gso_2025}
J.J.~Heckman, J.~McNamara, J.~Parra-Martinez and E.~Torres, \emph{{GSO} {Defects}: {IIA}/{IIB} {Walls} and the {Surprisingly} {Stable} \${\textbackslash}mathrm\{{R}\}7\$-{Brane}},  July, 2025.
\newblock 10.48550/arXiv.2507.21210.

\bibitem{hanany_orientifolds_2000}
A.~Hanany and B.~Kol, \emph{On {Orientifolds}, {Discrete} {Torsion}, {Branes} and {M} {Theory}}, \href{https://doi.org/10.1088/1126-6708/2000/06/013}{\emph{Journal of High Energy Physics} {\bfseries 2000} (2000) 013}.

\bibitem{dasgupta_orbifolds_1996}
K.~Dasgupta and S.~Mukhi, \emph{Orbifolds of {M}-theory}, \href{https://doi.org/10.1016/0550-3213(96)00070-3}{\emph{Nuclear Physics B} {\bfseries 465} (1996) 399}.

\bibitem{witten_five-branes_1996}
E.~Witten, \emph{Five-branes {And} \${M}\$-{Theory} {On} {An} {Orbifold}}, \href{https://doi.org/10.1016/0550-3213(96)00032-6}{\emph{Nuclear Physics B} {\bfseries 463} (1996) 383}.

\bibitem{freed_consistency_2021}
D.S.~Freed and M.J.~Hopkins, \emph{Consistency of {M}-{Theory} on nonorientable manifolds},  Jan., 2021.
\newblock 10.48550/arXiv.1908.09916.

\bibitem{AlvarezGarcia-heterotic-202x}
R.~Álvarez García, V.~Nevoa and S.~Raman, \emph{Heterotic branes, bordism, and fractional charge}, {\emph{To appear} (2026) }.

\bibitem{hsieh_anomaly_2019}
C.-T.~Hsieh, Y.~Tachikawa and K.~Yonekura, \emph{Anomaly of the {Electromagnetic} {Duality} of {Maxwell} {Theory}}, \href{https://doi.org/10.1103/PhysRevLett.123.161601}{\emph{Physical Review Letters} {\bfseries 123} (2019) 161601}.

\bibitem{aharony_s-folds_2016}
O.~Aharony, Y.~Tachikawa and K.~Gomi, \emph{S-folds and 4d {N}=3 superconformal field theories}, \href{https://doi.org/10.1007/JHEP06(2016)044}{\emph{Journal of High Energy Physics} {\bfseries 2016} (2016) 44}.

\bibitem{garcia-etxebarria_n3_2016}
I.~García-Etxebarria and D.~Regalado, \emph{N=3 four dimensional field theories}, \href{https://doi.org/10.1007/JHEP03(2016)083}{\emph{Journal of High Energy Physics} {\bfseries 2016} (2016) 83}.

\bibitem{witten_d-branes_1998}
E.~Witten, \emph{D-{Branes} {And} {K}-{Theory}}, \href{https://doi.org/10.1088/1126-6708/1998/12/019}{\emph{Journal of High Energy Physics} {\bfseries 1998} (1998) 019}.

\bibitem{kaidi_nonsupersymmetric_2023}
J.~Kaidi, K.~Ohmori, Y.~Tachikawa and K.~Yonekura, \emph{Nonsupersymmetric {Heterotic} {Branes}}, \href{https://doi.org/10.1103/PhysRevLett.131.121601}{\emph{Phys. Rev. Lett.} {\bfseries 131} (2023) 121601}.

\bibitem{freed_dirac_2001}
D.S.~Freed, \emph{Dirac {Charge} {Quantization} and {Generalized} {Differential} {Cohomology}},  Feb., 2001.
\newblock 10.48550/arXiv.hep-th/0011220.

\bibitem{polchinski_open_2006}
J.~Polchinski, \emph{Open {Heterotic} {Strings}}, \href{https://doi.org/10.1088/1126-6708/2006/09/082}{\emph{Journal of High Energy Physics} {\bfseries 2006} (2006) 082}.

\bibitem{bergshoeff_open_2006}
E.A.~Bergshoeff, G.W.~Gibbons and P.K.~Townsend, \emph{Open {M5}-branes}, \href{https://doi.org/10.1103/PhysRevLett.97.231601}{\emph{Physical Review Letters} {\bfseries 97} (2006) 231601}.

\bibitem{anderson_lectures_2023}
L.B.~Anderson, J.~Gray and M.~Larfors, \emph{Lectures on {Numerical} and {Machine} {Learning} {Methods} for {Approximating} {Ricci}-flat {Calabi}-{Yau} {Metrics}},  Dec., 2023.
\newblock 10.48550/arXiv.2312.17125.

\bibitem{larfors_numerical_2022}
M.~Larfors, A.~Lukas, F.~Ruehle and R.~Schneider, \emph{Numerical {Metrics} for {Complete} {Intersection} and {Kreuzer}-{Skarke} {Calabi}-{Yau} {Manifolds}},  May, 2022.
\newblock 10.48550/arXiv.2205.13408.

\bibitem{berglund_machine_2024}
P.~Berglund, G.~Butbaia, T.~Hübsch, V.~Jejjala, D.M.~Peña, C.~Mishra et~al., \emph{Machine {Learned} {Calabi}-{Yau} {Metrics} and {Curvature}}, \href{https://doi.org/10.4310/ATMP.2023.v27.n4.a3}{\emph{Advances in Theoretical and Mathematical Physics} {\bfseries 27} (2024) 1107}.

\bibitem{van_de_heisteeg_moduli-dependent_2022}
D.~van~de Heisteeg, C.~Vafa, M.~Wiesner and D.H.~Wu, \emph{Moduli-dependent {Species} {Scale}},  Dec., 2022.

\bibitem{van_de_heisteeg_species_2023}
D.~van~de Heisteeg, C.~Vafa, M.~Wiesner and D.H.~Wu, \emph{Species {Scale} in {Diverse} {Dimensions}}, .

\bibitem{raman_swampland_2024}
S.~Raman and C.~Vafa, \emph{Swampland and the {Geometry} of {Marked} {Moduli} {Spaces}},  June, 2024.
\newblock 10.48550/arXiv.2405.11611.

\bibitem{bedroya_holography_2025}
A.~Bedroya and P.J.~Steinhardt, \emph{Holography vs. {Scale} {Separation}},  Oct., 2025.
\newblock 10.48550/arXiv.2509.25313.

\bibitem{susskind_stretched_1993}
L.~Susskind, L.~Thorlacius and J.~Uglum, \emph{The {Stretched} {Horizon} and {Black} {Hole} {Complementarity}}, \href{https://doi.org/10.1103/PhysRevD.48.3743}{\emph{Physical Review D} {\bfseries 48} (1993) 3743}.

\end{thebibliography}\endgroup

\end{document}